# Molecular Multipoles and (Hyper)Polarizabilities from Buckingham Expansion: Revisited

Houxian Chen, Menglin Liu, Tianying Yan[*]

Institute of New Energy Material Chemistry, School of Materials Science and Engineering, National Institute for Advanced Materials, Nankai University, Tianjin 300350, China.

[*]corresponding author: tyan@nankai.edu.cn

**Abstract**

Buckingham expansion is important for understanding molecular multipoles and (hyper)polarizabilities. In this study, we give a complete derivation of Buckingham expansion in the traced form using successive Taylor series. Based on such derivation, a general Buckingham expansion in the traced form is proposed, and from which numerical calculations with finite field method of high accuracy can be achieved. The transformations from the traced multipoles and multipole-multipole polarizabilities to the corresponding traceless counterparts are realized with an auxiliary traced electric field gradient. The applications of the finite field method in this study show good agreements with previous theoretical calculations and experimental measurements.

## 1. Introduction

Multipoles and (hyper)polarizabilities are important molecular interactions[1, 2]. The interactions between molecule and external electric fields and field gradients, can be expanded with multipoles in external electric fields and field gradients, which is the well-known multipole expansion. Furthermore, if the molecule is polarizable, the multipoles are also inducible by the field and field gradients. The resultant interactions are expanded to include all the permanent and induced multipoles, with the induced components manifested in the (hyper)polarizabilities, and we denote in this study the resultant expression as Buckingham expansion[3-5]. Thus, as multiple external electric fields and field gradients $E$ are applied on the polarizable charge distribution $\rho(r)$, which can be expanded by permanent and induced multipoles $M(r)$, the responses via Buckingham expansion is shown schematically in Figure 1.

$$U = U_0 + q\varphi - \mu_\alpha E_\alpha - \frac{1}{2}\alpha_{\alpha\beta}E_\alpha E_\beta - \frac{1}{6}\beta_{\alpha\beta\gamma}E_\alpha E_\beta E_\gamma - \cdots$$
$$- \frac{1}{3}\Theta_{\alpha\beta}E_{\alpha\beta} - \frac{1}{3}A_{\gamma,\alpha\beta}E_\gamma E_{\alpha\beta} - \frac{1}{6}C_{\alpha\beta,\gamma\delta}E_{\alpha\beta}E_{\gamma\delta} - \cdots$$
$$- \frac{1}{15}\Omega_{\alpha\beta\gamma}E_{\alpha\beta\gamma} - \frac{1}{15}D_{\delta,\alpha\beta\gamma}E_\delta E_{\alpha\beta\gamma} - \cdots$$

**Figure 1.** A schematic Buckingham expansion, in which $E$ can be external electric fields and field gradients, $\rho(r)$ represents the polarizable charge distribution, and $M(r)$ denotes permanent and induced multipoles. The equation corresponds to Buckingham expansion in the traceless form, which is derived in Eq. (27).

Buckingham expansion is not only of interest theoretically, but also leads to important applications on determining molecular (hyper)polarizabilities[6] and quadrupole[3, 7]. By designing a four-wire electric field gradient condenser, which induces birefringence (Kerr effects) on the polar molecular system, inherently originated from the anisotropic molecular multipoles/(hyper)polarizabilities as represented by Buckingham expansion, Buckingham derived the method to detect the molecular quadrupole[3]. Such theoretical derivation was realized experimentally with a direct detection of the quadrupole of carbon dioxide[7].

Since it is generally difficult to detect molecular hyperpolarizabilities experimentally, except some small molecules of high symmetry, it is common nowadays to calculate molecular multipoles and (hyper)polarizabilities numerically against high level *ab initio* calculations, either using test charges to generate the desired finite external electric field and field gradients[8, 9], or manipulating the electric field tensors directly[10, 11]. Either way, the molecular multipoles and (hyper)polarizabilities can be obtained by finite difference with high accuracy up to $O(E^n)$, in which $E$ denotes the external fields or field gradients. In order to get numerical multipoles and (hyper)polarizabilities, especially for the multipole-multipole polarizabilities, accurately up to high-order $n$, complete Buckingham expansion needs to be known.

To the best of our knowledge, Buckingham expansion was originally proposed up to the fourth order of electric fields or field gradients in the traceless form, in which all the multipoles and (hyper)polarizabilities are traceless[3, 4]. For *ab initio* calculations, it is more convenient to work instead with the traced multipoles and (hyper)polarizabilities. Thus, a general Buckingham expansion in the traced form up to arbitrary order of electric fields or field gradients, as well as the corresponding transformation from traced to traceless multipoles and (hyper)polarizabilities, are desired. Aiming on filling the gaps, McLean and Yoshimine[12], Applequist[13], and Pedersen *et al.*[11] all derived the transformation from traced to traceless multipoles and (hyper)polarizabilities, in which the transformation of the multipole-multipole polarizabilities were achieved by the comparisons between Buckingham expansion in the traced and traceless forms. Considering the importance of Buckingham expansion, we believe that a re-visit of Buckingham expansion is needed in order to clarify some of the ambiguities. In this study, we give a complete derivation of Buckingham expansion, which we think is easily understood and more closely resembles the original form of Buckingham expansion, as we re-visit this important and interesting topic.

In this study, the derivation is based on solely cartesian tensors, because it is more relevant to the experimentally reported molecular multipoles and

(hyper)polarizabilities. The transformation between cartesian representation and spherical representation with solid harmonic spherical bases has been well documented[2, 11]. We derive in Sec. 2 Buckingham expansion in the traced form, using Taylor series, with necessary discussions on the Buckingham convention according to the original Buckingham expansion in the traceless form. We next propose an auxiliary traced field gradient, and from which to derive the Buckingham expansion in the traceless form in Sec. 3. By such method, we can easily obtain the transformation from traced multipoles and (hyper)polarizabilities to the corresponding traceless counterparts. While the former is well known, the latter may not be trivial without the auxiliary traced field gradient. Additionally, we demonstrate that the traceless features of multipoles as well as the multipole-multipole polarizabilities, are the consequence of the Laplace equation. In Sec. 4, we apply finite field method against high level *ab initio* calculations, utilizing Buckingham expansion, to numerically calculate the dipole polarizability, quadrupole, and quadrupole-quadrupole polarizability of carbon dioxide. In appendix A, we briefly summarize previous derivations, especially McLean and Yoshimine's derivation[12], and derive the conversions of the multipole-multipole polarizabilities of different conventions. In appendix B, we propose a general Buckingham expansion in the traced form, and derive the finite difference expressions of the traced quadrupole and quadrupole-quadrupole polarizability with accuracy of $O(E^4)$. Note that the derivation of any multipole or (hyper)polarizability is straight-forward using the finite field method in Appendix B, and the transformations from the traced quadrupole and quadrupole-quadrupole polarizability to the corresponding traceless counterparts are derived in Sec. 3. In appendix C, we propose a general formula for Buckingham expansion in the traceless form based on the discussions in the text.

**2. Buckingham expansion in the traced form.**

Considering the interaction of a charge distribution $\rho(r)$, centered at the origin $O$ of a cartesian coordinate system, with an external electrostatic perturbation, the external electrostatic potential at $r$ can be expanded via Taylor series around $O$, i.e.[2,

12],

$$\varphi(\bm{r}) = \varphi(0) - r_\alpha E_\alpha - \frac{1}{2} r_\alpha r_\beta E_{\alpha\beta} - \frac{1}{6} r_\alpha r_\beta r_\gamma E_{\alpha\beta\gamma} - \frac{1}{24} r_\alpha r_\beta r_\gamma r_\delta E_{\alpha\beta\gamma\delta} - \cdots \qquad (1)$$

in which $\varphi$, $E_\alpha$, $E_{\alpha\beta}$, $E_{\alpha\beta\gamma}$, and $E_{\alpha\beta\gamma\delta}$ denote external electric potential, field, gradients, *etc.*, and the Greek subscripts denote the cartesian coordinate component $\{x, y, z\}$. In Eq. (1), Einstein summation convention on the repeated subscripts is adopted throughout.

Using Eq. (1), the interaction between $\rho(\bm{r})$ and the externally perturbing $\varphi(\bm{r})$ is given by[2],

$$U - U_0 = \int_V dv \rho(\bm{r}) \varphi(\bm{r}) = q\varphi - \mu'_\alpha E_\alpha - \frac{1}{2} Q'_{\alpha\beta} E_{\alpha\beta} - \frac{1}{6} O'_{\alpha\beta\gamma} E_{\alpha\beta\gamma} - \frac{1}{24} H'_{\alpha\beta\gamma\delta} E_{\alpha\beta\gamma\delta} - \cdots \qquad (2)$$

in which $U_0$ is the energy without external field perturbations, and $q$, $\mu'_\alpha$, $Q'_{\alpha\beta}$, $O'_{\alpha\beta\gamma}$, and $H'_{\alpha\beta\gamma\delta}$ are monopole (charge), dipole, traced quadrupole, traced octopole, and traced hexadecapole, respectively, i.e.,

$$q = \int_V dv \rho(\bm{r}) \qquad (3a)$$

$$\mu'_\alpha = \int_V dv \rho(\bm{r}) r_\alpha \qquad (3b)$$

$$Q'_{\alpha\beta} = \int_V dv \rho(\bm{r}) r_\alpha r_\beta \qquad (3c)$$

$$O'_{\alpha\beta\gamma} = \int_V dv \rho(\bm{r}) r_\alpha r_\beta r_\gamma \qquad (3d)$$

$$H'_{\alpha\beta\gamma\delta} = \int_V dv \rho(\bm{r}) r_\alpha r_\beta r_\gamma r_\delta \qquad (3e)$$

$$\vdots$$

Eq. (3) stands because the external electrostatic perturbations, $\varphi$, $E_\alpha$, $E_{\alpha\beta}$, $E_{\alpha\beta\gamma}$, $E_{\alpha\beta\gamma\delta}$,

*etc.*, are independent of *r*, and the integration is performed over the entire space $V$ occupied by $\rho(r)$ which is well separated from the external source. The primes on the superscripts of the multipoles in Eqs. (2) and (3) remind that they may all be inducible if $\rho(r)$ is polarizable, i.e.,

$$\mu'_\alpha = \mu_\alpha + \mu_\alpha^{ind}, \quad Q'_{\alpha\beta} = Q_{\alpha\beta} + Q_{\alpha\beta}^{ind}, \quad O'_{\alpha\beta\gamma} = O_{\alpha\beta\gamma} + O_{\alpha\beta\gamma}^{ind}, \quad \cdots \tag{4}$$

in which $\mu_\alpha$, $Q_{\alpha\beta}$, and $O_{\alpha\beta\gamma}$ are the permanent multipoles, with the superscript "ind" in the second term on the RHS denotes the individually inducible counterpart. Such classification is subtle, because the inducible terms only appear with additional external perturbations. Otherwise, they are the permanent multipoles. Since there is no charge source or sink other than $\rho(r)$, $q$ is permanent and not inducible.

With additional external perturbation, the multipoles can be expanded, again, via Taylor series, i.e.,

$$\mu'_\alpha = -\frac{\partial U}{\partial E_\alpha} = \mu_\alpha + \alpha'_{\alpha\beta}E_\beta + \frac{1}{2}a'_{\alpha,\beta\gamma}E_{\beta\gamma} + \frac{1}{6}d'_{\alpha,\beta\gamma\delta}E_{\beta\gamma\delta} + \cdots \tag{5a}$$

$$Q'_{\alpha\beta} = -2\frac{\partial U}{\partial E_{\alpha\beta}} = Q_{\alpha\beta} + a'_{\gamma,\alpha\beta}E_\gamma + \frac{1}{2}\left(c'_{\alpha\beta,\gamma\delta} + c'_{\gamma\delta,\alpha\beta}\right)E_{\gamma\delta} + \frac{1}{6}g'_{\alpha\beta,\gamma\delta\eta}E_{\gamma\delta\eta} + \cdots$$
$$= Q_{\alpha\beta} + a'_{\gamma,\alpha\beta}E_\gamma + c'_{\alpha\beta,\gamma\delta}E_{\gamma\delta} + \frac{1}{6}g'_{\alpha\beta,\gamma\delta\eta}E_{\gamma\delta\eta} + \cdots \tag{5b}$$

$$O'_{\alpha\beta\gamma} = -6\frac{\partial U}{\partial E_{\alpha\beta\gamma}} = O_{\alpha\beta\gamma} + d'_{\delta,\alpha\beta\gamma}E_\delta + \frac{1}{2}g'_{\delta\eta,\alpha\beta\gamma}E_{\delta\eta} + \frac{1}{6}\left(p'_{\alpha\beta\gamma,\delta\eta\xi} + p'_{\delta\eta\xi,\alpha\beta\gamma}\right)E_{\delta\eta\xi} + \cdots$$
$$= O_{\alpha\beta\gamma} + d'_{\delta,\alpha\beta\gamma}E_\delta + \frac{1}{2}g'_{\delta\eta,\alpha\beta\gamma}E_{\delta\eta} + \frac{1}{3}p'_{\alpha\beta\gamma,\delta\eta\xi}E_{\delta\eta\xi} + \cdots \tag{5c}$$

$$H'_{\alpha\beta\gamma\delta} = -24\frac{\partial U}{\partial E_{\alpha\beta\gamma\delta}} = H_{\alpha\beta\gamma\delta} + \cdots \tag{5d}$$

in which all but the first term on the RHS are the induced multipoles by external electric field and gradients. $\alpha_{\alpha\beta}$ is dipole polarizability, $a_{\gamma,\alpha\beta}$ is dipole-quadrupole polarizability, $d_{\delta,\alpha\beta\gamma}$ is dipole-octopole polarizability, $c_{\alpha\beta,\gamma\delta} = c_{\gamma\delta,\alpha\beta}$ is quadrupole-quadrupole polarizability, $g_{\alpha\beta,\gamma\delta\eta}$ is quadrupole-octopole polarizability, and $p_{\alpha\beta\gamma,\delta\eta\xi} = p_{\delta\eta\xi,\alpha\beta\gamma}$ is octopole-octopole polarizability, respectively, which are

$$\alpha_{\alpha\beta} = \frac{\partial \mu_\alpha}{\partial E_\beta} = -\frac{\partial^2 U}{\partial E_\alpha \partial E_\beta} \tag{6a}$$

$$a_{\gamma,\alpha\beta} = 2\frac{\partial \mu_\gamma}{\partial E_{\alpha\beta}} = \frac{\partial Q_{\alpha\beta}}{\partial E_\gamma} = -2\frac{\partial^2 U}{\partial E_\gamma \partial E_{\alpha\beta}} \tag{6b}$$

$$d_{\delta,\alpha\beta\gamma} = 6\frac{\partial \mu_\delta}{\partial E_{\alpha\beta\gamma}} = \frac{\partial O_{\alpha\beta\gamma}}{\partial E_\delta} = -6\frac{\partial^2 U}{\partial E_\delta \partial E_{\alpha\beta\gamma}} \tag{6c}$$

$$c_{\alpha\beta,\gamma\delta} = \frac{\partial Q_{\alpha\beta}}{\partial E_{\gamma\delta}} = -2\frac{\partial^2 U}{\partial E_{\alpha\beta} \partial E_{\gamma\delta}} \tag{6d}$$

$$g_{\delta\eta,\alpha\beta\gamma} = 6\frac{\partial Q_{\delta\eta}}{\partial E_{\alpha\beta\gamma}} = 2\frac{\partial O_{\alpha\beta\gamma}}{\partial E_{\delta\eta}} = -12\frac{\partial^2 U}{\partial E_{\delta\eta} \partial E_{\alpha\beta\gamma}} \tag{6e}$$

$$p_{\alpha\beta\gamma,\delta\eta\xi} = 3\frac{\partial O_{\alpha\beta\gamma}}{\partial E_{\delta\eta\xi}} = -18\frac{\partial^2 U}{\partial E_{\alpha\beta\gamma} \partial E_{\delta\eta\xi}} \tag{6f}$$

in which the multipole-multipole polarizabilities are all in the traced form. The primes on the superscripts of the (hyper)polarizabilities in Eq. (5) remind that they may all be inducible upon additional external perturbation. For example,

$$\alpha'_{\alpha\beta} = \alpha_{\alpha\beta} + \beta'_{\alpha\beta\gamma}E_\gamma + \frac{1}{2}b'_{\alpha\beta,\gamma\delta}E_{\gamma\delta} + \cdots \tag{7a}$$

$$\beta'_{\alpha\beta\gamma} = \beta_{\alpha\beta\gamma} + \gamma'_{\alpha\beta\gamma\delta}E_\delta + \cdots \tag{7b}$$

$$\begin{aligned} a'_{\gamma,\alpha\beta} &= a_{\gamma,\alpha\beta} + b'_{\gamma\delta,\alpha\beta}E_\delta + \frac{1}{2}\left(c^{(2)'}_{\gamma,\alpha\beta,\delta\eta} + c^{(2)'}_{\gamma,\delta\eta,\alpha\beta}\right)E_{\delta\eta} + \cdots \\ &= a_{\gamma,\alpha\beta} + b'_{\gamma\delta,\alpha\beta}E_\delta + c^{(2)'}_{\gamma,\alpha\beta,\delta\eta}E_{\delta\eta} + \cdots \end{aligned} \tag{7c}$$

in which $\beta_{\alpha\beta\gamma}$ and $\gamma_{\alpha\beta\gamma\delta}$ are the first and second dipole hyperpolarizability, respectively, $b_{\alpha\beta,\gamma\delta}$ is dipole-dipole-quadrupole polarizability, and $c^{(2)}_{\gamma,\alpha\beta,\delta\eta} = c^{(2)}_{\gamma,\delta\eta,\alpha\beta}$ is

dipole-quadrupole-quadrupole polarizability, which are

$$\beta_{\alpha\beta\gamma} = \frac{\partial \alpha_{\alpha\beta}}{\partial E_\gamma} = -\frac{\partial^3 U}{\partial E_\alpha \partial E_\beta \partial E_\gamma} \tag{8a}$$

$$\gamma_{\alpha\beta\gamma\delta} = \frac{\partial \beta_{\alpha\beta\gamma}}{\partial E_\delta} = -\frac{\partial^4 U}{\partial E_\alpha \partial E_\beta \partial E_\gamma \partial E_\delta} \tag{8b}$$

$$b_{\gamma\delta,\alpha\beta} = 2\frac{\partial \alpha_{\gamma\delta}}{\partial E_{\alpha\beta}} = \frac{\partial a_{\gamma,\alpha\beta}}{\partial E_\delta} = -2\frac{\partial^3 U}{\partial E_\gamma \partial E_\delta \partial E_{\alpha\beta}} \tag{8c}$$

$$c^{(2)}_{\gamma,\alpha\beta,\delta\eta} = \frac{\partial a_{\gamma,\alpha\beta}}{\partial E_{\delta\eta}} = -2\frac{\partial^3 U}{\partial E_\gamma \partial E_{\alpha\beta} \partial E_{\delta\eta}} \tag{8d}$$

In the Taylor series in Eqs. (5) and (7), the Greek subscripts of the multipole-multipole polarizabilities are grouped, with different groups separated by comma. The convention in this study is that the subscripts in the same group are permutable, while the exchange of the groups of the subscripts separated by comma is not allowed. Thus, the singly grouped subscripts of electric fields, gradients, multipoles, dipole (hyper)polarizabilities are permutable. For the multiply grouped subscripts, taking dipole-quadrupole polarizability as example, $a_{\gamma,\alpha\beta}$ and $a_{\gamma,\beta\alpha}$ are allowed in the summation of Eqs. (5a–5b), but $a_{\alpha\beta,\gamma}$ and $a_{\beta\alpha,\gamma}$ are not allowed, though the four terms are all equal. Thus, such non-exchangeable group convention includes totally 27 $a_{\gamma,\alpha\beta}$ terms, while allowing group exchange of $a_{\alpha\beta,\gamma}$ merely double the number of terms, with the values of the terms of the former 2 times the corresponding terms of the latter. Similar argument also applies on $b_{\gamma\delta,\alpha\beta}$, $d_{\delta,\alpha\beta\gamma}$, $g_{\delta\eta,\alpha\beta\gamma}$, etc.. On the other hand, it is notable that the groups of the subscripts of $c_{\alpha\beta,\gamma\delta}$ and $p_{\alpha\beta\gamma,\delta\eta\xi}$, with same number of subscripts in some groups, are inherently exchangeable. Thus, they are explicitly included in Eqs. (5a–5b). The same argument also accounts for the $c^{(2)}_{\gamma,\alpha\beta,\delta\eta}$ in Eq. (7c), because the groups, $\alpha\beta$ and $\delta\eta$, in the subscripts of $c^{(2)}_{\gamma,\alpha\beta,\delta\eta}$ are inherently exchangeable. Generally, if there are $n$ inherently exchangeable groups in the subscripts of a multipole-multipole polarizability, the factor on the corresponding term is *n*! to account for the permutations of the inherently exchangeable groups.

It should be noted that the above convention was implicitly adopted by Buckingham expansion in its originally proposed form[3-5], and we thus call it Buckingham convention hereafter. Other conventions are also adopted in literatures[12-14]. Since Buckingham expansion has been widely adopted in community on reporting the multipole-multipole polarizabilities[8, 9, 11], other conventions are often converted to Buckingham convention. One conversion between two different conventions is given in Appendix A. Moreover, Buckingham convention leads to a simple generalized Buckingham expansion in the traced form, as shown in Appendix B.

Though the above Taylor series can be continued toward infinite terms, we include hopefully in Eqs. (5–8) most of the important ones. The induced multipoles and (hyper)polarizabilities can then be obtained via successive integrations in a reversible manner[13]. For example, the induced component of the dipole polarizability $\alpha_{\alpha\beta}$ on external fields $E_\gamma$ and $E_\delta$, from Eq. (7b), is

$$\alpha'_{\alpha\beta} - \alpha_{\alpha\beta} = \int_0^{E_\gamma} \beta'_{\alpha\beta\gamma} \mathrm{d}(\lambda E_\gamma) = \int_0^1 \left(\beta_{\alpha\beta\gamma} + \gamma'_{\alpha\beta\gamma\delta}\lambda E_\delta + \cdots\right) E_\gamma \mathrm{d}\lambda$$
$$= \beta_{\alpha\beta\gamma} E_\gamma + \frac{1}{2}\gamma'_{\alpha\beta\gamma\delta} E_\gamma E_\delta + \cdots \tag{9}$$

For the second dipole hyperpolarizability $\gamma_{\alpha\beta\gamma\delta}$ in the final expression in Eq. (9), the factor 1/2 arises from the integration over the parameter $\lambda(0 \le \lambda \le 1)$ for a reversible polarization process[13]. Feeding the induction component of $\alpha_{\alpha\beta}$ in Eq. (9) back to Eq. (7a), we have

$$\alpha'_{\alpha\beta} = \alpha_{\alpha\beta} + \beta_{\alpha\beta\gamma} E_\gamma + \frac{1}{2}\gamma'_{\alpha\beta\gamma\delta} E_\gamma E_\delta + \frac{1}{2}b'_{\alpha\beta,\gamma\delta} E_{\gamma\delta} + \cdots \tag{7a'}$$

which gives the differential dipole polarizability[6] $\alpha_{\alpha\beta}$ on external $E_\gamma$, $E_\delta$, and $E_{\gamma\delta}$. Integrating Eq. (7a′) over $E_\beta$ and Eq. (7c) over $E_{\beta\gamma}$ with the reversible polarization process, respectively, and avoiding the double counting on the term involving dipole-dipole-quadrupole polarizability $b_{\alpha\beta,\gamma\delta}$, we can obtain the induced dipole $\boldsymbol{\mu}^{\text{ind}}$, i.e.,

$$\begin{aligned}
\mu_\alpha^{\text{ind}} &= \int_0^1 \left( \alpha_{\alpha\beta} + \beta_{\alpha\beta\gamma}\lambda E_\gamma + \frac{1}{2}\gamma'_{\alpha\beta\gamma\delta}\lambda E_\gamma \lambda E_\delta + \cdots \right) E_\beta \mathrm{d}\lambda \\
&+ \frac{1}{2}\int_0^1 \left( a_{\alpha,\beta\gamma} + b'_{\alpha\delta,\beta\gamma}E_\delta + c^{(2)'}_{\alpha,\beta\gamma,\delta\eta}\lambda E_{\delta\eta} + \cdots \right) E_{\beta\gamma} \mathrm{d}\lambda \\
&= \alpha_{\alpha\beta}E_\beta + \frac{1}{2}\beta_{\alpha\beta\gamma}E_\beta E_\gamma + \frac{1}{6}\gamma'_{\alpha\beta\gamma\delta}E_\beta E_\gamma E_\delta + \cdots \\
&+ \frac{1}{2}a_{\alpha,\beta\gamma}E_{\beta\gamma} + \frac{1}{2}b'_{\alpha\delta,\beta\gamma}E_\delta E_{\beta\gamma} + \frac{1}{4}c^{(2)'}_{\alpha,\beta\gamma,\delta\eta}E_{\beta\gamma}E_{\delta\eta} + \cdots
\end{aligned} \quad (10\text{a})$$

$$\begin{aligned}
Q_{\alpha\beta}^{\text{ind}} &= \int_0^1 \left( a_{\gamma,\alpha\beta} + b'_{\gamma\delta,\alpha\beta}\lambda E_\delta + c^{(2)'}_{\gamma,\alpha\beta,\delta\eta}E_{\delta\eta} + \cdots \right) E_\gamma \mathrm{d}\lambda \\
&= a_{\gamma,\alpha\beta}E_\gamma + \frac{1}{2}b'_{\gamma\delta,\alpha\beta}E_\gamma E_\delta + c^{(2)'}_{\gamma,\alpha\beta,\delta\eta}E_\gamma E_{\delta\eta} + \cdots
\end{aligned} \quad (10\text{b})$$

Feeding Eq. (10a) back to Eq. (5a), and Eq. (10b) to Eq. (5b), we have

$$\begin{aligned}
\mu'_\alpha &= \mu_\alpha + \alpha_{\alpha\beta}E_\beta + \frac{1}{2}\beta_{\alpha\beta\gamma}E_\beta E_\gamma + \frac{1}{6}\gamma'_{\alpha\beta\gamma\delta}E_\beta E_\gamma E_\delta + \cdots \\
&+ \frac{1}{2}a_{\alpha,\beta\gamma}E_{\beta\gamma} + \frac{1}{2}b'_{\alpha\delta,\beta\gamma}E_\delta E_{\beta\gamma} + \frac{1}{4}c^{(2)'}_{\alpha,\beta\gamma,\delta\eta}E_{\beta\gamma}E_{\delta\eta} + \cdots \\
&+ \frac{1}{6}d'_{\alpha,\beta\gamma\delta}E_{\beta\gamma\delta} + \cdots
\end{aligned} \quad (5\text{a}')$$

$$\begin{aligned}
Q'_{\alpha\beta} &= Q_{\alpha\beta} + a_{\gamma,\alpha\beta}E_\gamma + \frac{1}{2}b'_{\gamma\delta,\alpha\beta}E_\gamma E_\delta + c^{(2)'}_{\gamma,\alpha\beta,\delta\eta}E_\gamma E_{\delta\eta} + \cdots \\
&+ c'_{\alpha\beta,\gamma\delta}E_{\gamma\delta} + \frac{1}{6}g'_{\alpha\beta,\gamma\delta\eta}E_{\gamma\delta\eta} + \cdots
\end{aligned} \quad (5\text{b}')$$

The Buckingham expansion in the traced form can be obtained via successive integration of Eqs. (5a–5d), with Eqs. (5a–5b) replaced by Eqs. (5a′–5b′). Thus,

$$U - U_0 = -\int_0^1 \mu'_\alpha E_\alpha d\lambda = -\left( \mu_\alpha + \frac{1}{2}\alpha_{\alpha\beta}E_\beta + \frac{1}{3!}\beta_{\alpha\beta\gamma}E_\beta E_\gamma + \frac{1}{4!}\gamma_{\alpha\beta\gamma\delta}E_\beta E_\gamma E_\delta \right.$$
$$\left. + \frac{1}{2}a_{\alpha,\beta\gamma}E_{\beta\gamma} + \frac{1}{4}b_{\alpha\beta,\gamma\delta}E_\beta E_{\gamma\delta} + \frac{1}{4}c^{(2)}_{\alpha,\beta\gamma,\delta\eta}E_{\beta\gamma}E_{\delta\eta} + \frac{1}{3!}d_{\alpha,\beta\gamma\delta}E_{\beta\gamma\delta} + \cdots \right) E_\alpha \quad (11a)$$

$$U - U_0 = -\frac{1}{2}\int_0^1 Q'_{\alpha\beta} E_{\alpha\beta} d\lambda$$
$$= -\frac{1}{2}\left( Q_{\alpha\beta} + a_{\gamma,\alpha\beta}E_\gamma + \frac{1}{2}b_{\gamma\delta,\alpha\beta}E_\gamma E_\delta + \frac{1}{2}c^{(2)}_{\gamma,\alpha\beta,\delta\eta}E_\gamma E_{\delta\eta} + \frac{1}{2}c_{\alpha\beta,\gamma\delta}E_{\gamma\delta} + \frac{1}{3!}g_{\alpha\beta,\gamma\delta\eta}E_{\gamma\delta\eta} + \cdots \right) E_{\alpha\beta}$$

(11b)

$$U - U_0 = -\frac{1}{6}\int_0^1 O'_{\alpha\beta\gamma} E_{\alpha\beta\gamma} d\lambda$$
$$= -\frac{1}{6}\left( O_{\alpha\beta\gamma} + d_{\delta,\alpha\beta\gamma}E_\delta + \frac{1}{2}g_{\delta\eta,\alpha\beta\gamma}E_{\delta\eta} + \frac{1}{6}p_{\alpha\beta\gamma,\delta\eta\xi}E_{\delta\eta\xi} + \cdots \right) E_{\alpha\beta\gamma} \quad (11c)$$

$$U - U_0 = -\frac{1}{24}\int_0^1 H'_{\alpha\beta\gamma\delta} E_{\alpha\beta\gamma\delta} d\lambda = -\frac{1}{24}\left( H_{\alpha\beta\gamma\delta} + \cdots \right) E_{\alpha\beta\gamma\delta} \quad (11d)$$

$$\vdots$$

in which the primes on the superscript are dropped in the final expression, because they are the permanent ones without additional external electric fields or gradients other than the listed in the otherwise presented ones. With such understanding, the superscript primes are dropped hereafter. Using Eq. (11), Eq. (2) can be further expanded to include both multipoles as well as (hyper)polarizabilities. Summing Eq. (11), and skipping the same terms to avoid double count, we can expand the energy of a polarizable charge distribution. The resultant expansion is the Buckingham expansion in the traced form, i.e.,

$$U = U_0 + q\varphi$$
$$- \left( \mu_\alpha E_\alpha + \frac{1}{2} \alpha_{\alpha\beta} E_\alpha E_\beta + \frac{1}{6} \beta_{\alpha\beta\gamma} E_\alpha E_\beta E_\gamma + \frac{1}{24} \gamma_{\alpha\beta\gamma\delta} E_\alpha E_\beta E_\gamma E_\delta + \cdots \right)$$
$$- \frac{1}{2} \left[ \left( Q_{\alpha\beta} + a_{\gamma,\alpha\beta} E_\gamma + \frac{1}{2} b_{\gamma\delta,\alpha\beta} E_\gamma E_\delta + \cdots \right) + \frac{1}{2} \left( c_{\alpha\beta,\gamma\delta} + c^{(2)}_{\eta,\alpha\beta,\gamma\delta} E_\eta + \cdots \right) E_{\gamma\delta} + \cdots \right] E_{\alpha\beta} \quad (12)$$
$$- \frac{1}{6} \left( O_{\alpha\beta\gamma} + d_{\delta,\alpha\beta\gamma} E_\delta + \frac{1}{2} g_{\delta\eta,\alpha\beta\gamma} E_{\delta\eta} + \frac{1}{6} p_{\delta\eta\xi,\alpha\beta\gamma} E_{\delta\eta\xi} + \cdots \right) E_{\alpha\beta\gamma}$$
$$- \frac{1}{24} H_{\alpha\beta\gamma\delta} E_{\alpha\beta\gamma\delta} - \cdots$$

From Eq. (12), we can deduce that for a polarizable charge distribution the Buckingham expansion in the traced form can be constructed via successive Taylor series toward infinite terms, and all the multipoles in Eq. (4) and the multipole-multipole polarizabilities in Eqs. (6) and (8) are the traced ones. Note that Buckingham convention is adopted throughout the derivation. In Appendix B, we give the general formula for the complete Buckingham expansion in the traced form. In the actual numerical calculations, it is convenient to get the traced multipoles and polarizabilities, and then transform them to the traceless counterparts if needed, as discussed below.

**3. Buckingham Expansion in the traceless form.**

Without losing generality, we consider the externally perturbed $\varphi$ is generated by a test positive unit electron charge $e$ at $\mathbf{R}$ far from $O$ on which the charge distribution $\rho(\mathbf{r})$ is located. The electrostatic potential at $\mathbf{r}$ due to $e$ at $\mathbf{R}$ is $\varphi(\mathbf{r}) = e/|\mathbf{R} - \mathbf{r}|$, which, upon binomial expansion with respect to $O$, is

$$\varphi(\mathbf{r}) = e \left( \frac{1}{R} + r_\alpha \frac{\partial R^{-1}}{\partial R_\alpha} - \frac{1}{2} r_\alpha r_\beta \frac{\partial^2 R^{-1}}{\partial R_\alpha \partial R_\beta} + \frac{1}{3!} r_\alpha r_\beta r_\gamma \frac{\partial^3 R^{-1}}{\partial R_\alpha \partial R_\beta \partial R_\gamma} - \frac{1}{4!} r_\alpha r_\beta r_\gamma r_\delta \frac{\partial^4 R^{-1}}{\partial R_\alpha \partial R_\beta \partial R_\gamma \partial R_\delta} + \cdots \right)$$
(13)

Comparing Eq. (13) with Eq. (1), we have the external potential, electric field, and gradients at *O*, due to *e* at **R**, i.e.,

$$\varphi(0) = \frac{e}{R} \tag{14a}$$

$$E_\alpha = -e\frac{\partial R^{-1}}{\partial R_\alpha} = e\frac{R_\alpha}{R^3} \tag{14b}$$

$$E_{\alpha\beta} = e\frac{\partial^2 R^{-1}}{\partial R_\alpha \partial R_\beta} = e\frac{1}{R^5}\left(3R_\alpha R_\beta - R^2 \delta_{\alpha\beta}\right) \tag{14c}$$

$$E_{\alpha\beta\gamma} = -e\frac{\partial^3 R^{-1}}{\partial R_\alpha \partial R_\beta \partial R_\gamma} = e\frac{1}{R^7}\left[15R_\alpha R_\beta R_\gamma - 3R^2\left(R_\alpha \delta_{\beta\gamma} + R_\beta \delta_{\alpha\gamma} + R_\gamma \delta_{\alpha\beta}\right)\right] \tag{14d}$$

$$\begin{aligned}E_{\alpha\beta\gamma\delta} = e\frac{\partial^4 R^{-1}}{\partial R_\alpha \partial R_\beta \partial R_\gamma \partial R_\delta} = e\frac{1}{R^9}\Big[&105 R_\alpha R_\beta R_\gamma R_\delta \\ -15R^2\big(&R_\alpha R_\beta \delta_{\gamma\delta} + R_\alpha R_\gamma \delta_{\beta\delta} + R_\alpha R_\delta \delta_{\beta\gamma} + R_\beta R_\gamma \delta_{\alpha\delta} + R_\beta R_\delta \delta_{\alpha\gamma} + R_\gamma R_\delta \delta_{\alpha\beta}\big) \\ +3R^4\big(&\delta_{\alpha\beta}\delta_{\gamma\delta} + \delta_{\alpha\gamma}\delta_{\beta\delta} + \delta_{\alpha\delta}\delta_{\beta\gamma}\big)\Big]\end{aligned} \tag{14e}$$

$\vdots$

in which $\delta_{\alpha\beta}$ denotes Kronecker symbol, which is 1 if *α* = *β* and is 0 otherwise. The subscripts *α*, *β*, *γ*, *δ* on the LHS of Eq. (14) refer to a specific cartesian component of electric field or gradients.

From Eq. (14), it is evident that the electric field gradients are symmetric with respect to the permutations of the subscripts. Also, they are traceless, i.e., $E_{\alpha\alpha} = E_{\alpha\beta\beta} = E_{\alpha\beta\gamma\gamma} = 0$, *etc.*, with repeated dummy variable summing over $\{x, y, z\}$, because *O* and **R** are spatially well separated and Laplace equation is satisfied on any place other than **R** where *e* is located[2, 4, 12]. We next define auxiliary external electric field gradients in the traced form, $F_{\alpha\beta}$, $F_{\alpha\beta\gamma}$, $F_{\alpha\beta\gamma\delta}$, with the aid from Eqs. (14c–14e), i.e.,

$$F_{\alpha\beta} = e\frac{3R_\alpha R_\beta}{R^5} \tag{15a}$$

$$F_{\alpha\beta\gamma} = e\frac{15 R_\alpha R_\beta R_\gamma}{R^7} \tag{15b}$$

$$F_{\alpha\beta\gamma\delta} = e\frac{105 R_\alpha R_\beta R_\gamma R_\delta}{R^9} \tag{15c}$$

Since electric field $E_\alpha$ is vector, it does not possess a "trace". Using Eq. (15), the electric field gradients in Eqs. (14c–14e) can be written as

$$E_{\alpha\beta} = F_{\alpha\beta} - \frac{1}{3} F_{\mu\mu} \delta_{\alpha\beta} \tag{16a}$$

$$E_{\alpha\beta\gamma} = F_{\alpha\beta\gamma} - \frac{1}{5}\left( F_{\alpha\mu\mu}\delta_{\beta\gamma} + F_{\mu\beta\mu}\delta_{\alpha\gamma} + F_{\mu\mu\gamma}\delta_{\alpha\beta} \right) \tag{16b}$$

$$\begin{aligned}
E_{\alpha\beta\gamma\delta} &= F_{\alpha\beta\gamma\delta} - \frac{1}{7}\left( F_{\alpha\beta\mu\mu}\delta_{\gamma\delta} + F_{\alpha\mu\gamma\mu}\delta_{\beta\delta} + F_{\alpha\mu\mu\delta}\delta_{\beta\gamma} + F_{\mu\beta\gamma\mu}\delta_{\alpha\delta} + F_{\mu\beta\mu\delta}\delta_{\alpha\gamma} + F_{\mu\mu\gamma\delta}\delta_{\alpha\beta} \right) \\
&+ \frac{1}{35}\left( F_{\mu\mu\nu\nu}\delta_{\alpha\beta}\delta_{\gamma\delta} + F_{\mu\nu\mu\nu}\delta_{\alpha\gamma}\delta_{\beta\delta} + F_{\mu\nu\nu\mu}\delta_{\alpha\delta}\delta_{\beta\gamma} \right)
\end{aligned} \tag{16c}$$

in which the places of the subscripts $\alpha$, $\beta$, $\gamma$, $\delta$ on the RHS are consistent with those on the LHS, which represents a specific cartesian component of electric field or gradient. Eq. (16) is derived by using the identity $R^2 = R_\mu R_\mu = R_x^2 + R_y^2 + R_z^2$, since the subscript $\mu \in \{x, y, z\}$ is dummy, and similarly $R^4 = R_\mu R_\mu R_\nu R_\nu = R_\mu R_\nu R_\mu R_\nu = R_\mu R_\nu R_\nu R_\mu$, in which the subscripts $\mu$ and $\nu$ are both dummy. It is easy to verify the traceless feature of electric field gradients by either using Eqs. (14c–14e), or, equivalently, using Eq. (16), i.e.,

$$E_{\alpha\alpha} = 0 \tag{17a}$$

$$E_{\alpha\alpha\gamma} = E_{\alpha\beta\alpha} = E_{\alpha\beta\beta} = 0 \tag{17b}$$

$$E_{\alpha\alpha\gamma\delta} = E_{\alpha\beta\alpha\delta} = E_{\alpha\beta\gamma\alpha} = E_{\alpha\beta\beta\delta} = E_{\alpha\beta\gamma\beta} = E_{\alpha\beta\gamma\gamma} = 0 \tag{17c}$$

For the above identities, Eq. (17a) is obtained via Eq. (16a), i.e., $E_{\alpha\alpha} = F_{\alpha\alpha} - \frac{1}{3}F_{\eta\eta}\delta_{\alpha\alpha} = 0$, using the identity $\delta_{\alpha\alpha}$ for the repeated dummy subscript $\alpha$.

Using the first identity $E_{\alpha\alpha\gamma}$ in Eq. (17b) as example, we have from Eq. (16b)

$$E_{\alpha\alpha\gamma} = F_{\alpha\alpha\gamma} - \frac{1}{5}\left(F_{\alpha\mu\mu}\delta_{\alpha\gamma} + F_{\mu\alpha\mu}\delta_{\alpha\gamma} + F_{\mu\mu\gamma}\delta_{\alpha\alpha}\right) = F_{\alpha\alpha\gamma} - \frac{1}{5}\left(F_{\gamma\mu\mu} + F_{\mu\gamma\mu} + 3F_{\mu\mu\gamma}\right) = 0$$

in which we use the fact that the repeated subscript $\alpha$ is dummy, so that $\delta_{\alpha\alpha} = 3$ and $\alpha$ can be replaced by another dummy variable $\mu$, i.e., $F_{\alpha\alpha\gamma} = F_{\mu\mu\gamma}$. Additionally, $F_{\gamma\mu\mu} = F_{\mu\gamma\mu} = F_{\mu\mu\gamma}$ since $F_{\alpha\beta\gamma}$ is invariant upon the permutation of the subscripts. Similar arguments apply on the other identities in Eqs. (17b–17c).

The traceless feature of the electric field gradients imposes the traceless condition on the corresponding multipoles. Taking, again, $E_{\alpha\beta}$ as example, and feeding $E_{\alpha\beta}$ in Eq. (16a) to the corresponding traced quadrupole $Q_{\alpha\beta}$ contribution to the energy in Eq. (2), we have $-\frac{1}{2}Q_{\alpha\beta}E_{\alpha\beta} = -\frac{1}{2}Q_{\alpha\beta}\left(F_{\alpha\beta} - \frac{1}{3}F_{\mu\mu}\delta_{\alpha\beta}\right) = -\frac{1}{2}\left(Q_{\alpha\beta}F_{\alpha\beta} - \frac{1}{3}Q_{\alpha\alpha}F_{\mu\mu}\right)$. Since all the subscripts are dummy, the last term can be written as $Q_{\alpha\alpha}F_{\mu\mu} = Q_{\mu\mu}F_{\alpha\alpha} = Q_{\mu\mu}\delta_{\alpha\beta}F_{\alpha\beta}$. Thus,

$$-\frac{1}{2}Q_{\alpha\beta}E_{\alpha\beta} = -\frac{1}{2}\left(Q_{\alpha\beta}F_{\alpha\beta} - \frac{1}{3}Q_{\mu\mu}\delta_{\alpha\beta}F_{\alpha\beta}\right) = -\frac{1}{2}\left(Q_{\alpha\beta} - \frac{1}{3}Q_{\mu\mu}\delta_{\alpha\beta}\right)F_{\alpha\beta} = -\frac{1}{2}\theta_{\alpha\beta}F_{\alpha\beta}$$

(18)

It is obvious that $\theta_{\alpha\beta} = Q_{\alpha\beta} - \frac{1}{3}Q_{\mu\mu}\delta_{\alpha\beta}$ on the RHS of Eq. (18) is traceless, i.e., $\theta_{\alpha\alpha} = 0$, for the same reasoning in Eq. (17). Thus, $\theta_{\alpha\beta}$ can be defined as the traceless quadrupole. On the other hand, such definition is not unique, because the traceless feature is maintained upon multiplication of any factor $f_\Theta$, i.e., $f_\Theta\theta_{\alpha\beta}$. Gaussian package[15] apparently adopts $f_\Theta = 1$, while in nuclear physics it adopts $f_\Theta = 3$[16]. Here, we adopt Buckingham's definition with $f_\Theta = 3/2$, so that the traceless quadrupole is defined to be[4]

$$\Theta_{\alpha\beta} = f_\Theta \theta_{\alpha\beta} = \frac{1}{2}\left(3Q_{\alpha\beta} - Q_{\mu\mu}\delta_{\alpha\beta}\right) \tag{19}$$

Thus, the traced $Q_{\alpha\beta}$ in Eq. (3c) can be transformed to the traceless $\Theta_{\alpha\beta}$ with the above equation. Using Eq. (19) with $f_\Theta = 3/2$, the energy $-\frac{1}{2}Q_{\alpha\beta}E_{\alpha\beta}$, in the presence of solely external electric field gradient, needs to be scaled by $\frac{1}{f_\Theta} = \frac{2}{3}$. Using Eqs. (11b), (18), and (19), we have

$$U - U_0 = -\frac{1}{2}Q_{\alpha\beta}E_{\alpha\beta} = -\frac{1}{2}\frac{1}{f_\Theta}f_\Theta\theta_{\alpha\beta}F_{\alpha\beta} = -\frac{1}{3}\Theta_{\alpha\beta}F_{\alpha\beta} = -\frac{1}{3}\Theta_{\alpha\beta}E_{\alpha\beta} \tag{20}$$

The validation of last expression of Eq. (20) is due to the traceless feature of $E_{\alpha\beta}$ and $\Theta_{\alpha\beta}$. Using Eq. (16a) and identity $\Theta_{\alpha\alpha} = 0$, the energy in external electric field gradient is

$$-\frac{1}{3}\Theta_{\alpha\beta}E_{\alpha\beta} = -\frac{1}{3}\Theta_{\alpha\beta}\left(F_{\alpha\beta} - \frac{1}{3}F_{\mu\mu}\delta_{\alpha\beta}\right) = -\frac{1}{3}\Theta_{\alpha\beta}F_{\alpha\beta} + \frac{1}{9}\Theta_{\alpha\alpha}F_{\mu\mu} = -\frac{1}{3}\Theta_{\alpha\beta}F_{\alpha\beta}$$

The above relation validates the last expression of Eq. (20).

From Eq. (20), we can see that it is equivalent to use traced $Q_{\alpha\beta}$ with traceless $E_{\alpha\beta}$ in Eq. (14c) or Eq. (16a), or to use traceless $\Theta_{\alpha\beta}$, with the auxiliary traced $F_{\alpha\beta}$ in Eq. (15a). On the other hand, using both traceless multipole and traceless electric field gradient is redundant. With such understanding, from $Q_{\alpha\beta} = -2\frac{\partial U}{\partial E_{\alpha\beta}}$, we see that when the charge distribution $\rho(\mathbf{r})$ senses a variation of the traceless $E_{\alpha\beta}$, it feeds back the traced $Q_{\alpha\beta}$ in response. In contrast, Eq. (20) also tells that $\Theta_{\alpha\beta} = -3\frac{\partial U}{\partial F_{\alpha\beta}}$. Thus, when the charge distribution $\rho(\mathbf{r})$ senses a variation of the traced $F_{\alpha\beta}$, it feeds back the traceless $\Theta_{\alpha\beta}$ in response. Using Eq. (16a), we have

$$J_{\gamma\delta,\alpha\beta} = \frac{\partial E_{\gamma\delta}}{\partial F_{\alpha\beta}} = \frac{\partial F_{\gamma\delta}}{\partial F_{\alpha\beta}} - \frac{1}{3}\frac{\partial F_{\mu\mu}}{\partial F_{\alpha\beta}}\delta_{\gamma\delta} = \delta_{\alpha\gamma}\delta_{\beta\delta} - \frac{1}{3}\delta_{\alpha\beta}\delta_{\gamma\delta} \qquad (21a)$$

in which $\frac{\partial F_{\mu\mu}}{\partial F_{\alpha\beta}} = \delta_{\mu\alpha}\delta_{\mu\beta} = \delta_{\alpha\beta}$ is used. The traceless quadrupole, $\Theta_{\alpha\beta}$, can then be calculated, equivalently, via

$$\Theta_{\alpha\beta} = -f_\Theta 2\frac{\partial U}{\partial F_{\alpha\beta}} = -\frac{3}{2}2\frac{\partial U}{\partial E_{\gamma\delta}}\frac{\partial E_{\gamma\delta}}{\partial F_{\alpha\beta}} = \frac{3}{2}Q_{\gamma\delta}\left(\delta_{\alpha\gamma}\delta_{\beta\delta} - \frac{1}{3}\delta_{\alpha\beta}\delta_{\gamma\delta}\right) = \frac{1}{2}\left(3Q_{\alpha\beta} - Q_{\mu\mu}\delta_{\alpha\beta}\right)$$

(22a)

in which $Q_{\alpha\beta} = -2\frac{\partial U}{\partial E_{\alpha\beta}}$ as in Eq. (5b), and $f_\Theta = 3/2$ as suggested by Buckingham are adopted. Thus, Eq. (22a) gives the same traceless $\Theta_{\alpha\beta}$ in Eq. (19), as expected.

The differentiation of energy with respect to the auxiliary traced electric field gradient in Eq. (16) makes it simple and consistent for the transformation from the traced multipoles and multipole-multipole polarizabilities to the corresponding traceless counterparts. From Eqs. (16b–16c), we have

$$J_{\delta\eta\xi,\alpha\beta\gamma} = \frac{\partial E_{\delta\eta\xi}}{\partial F_{\alpha\beta\gamma}} = \delta_{\alpha\delta}\delta_{\beta\eta}\delta_{\gamma\xi} - \frac{1}{5}\left(\delta_{\delta\alpha}\delta_{\beta\gamma}\delta_{\eta\xi} + \delta_{\beta\eta}\delta_{\alpha\gamma}\delta_{\delta\xi} + \delta_{\gamma\xi}\delta_{\alpha\beta}\delta_{\delta\eta}\right) \qquad (21b)$$

$$J_{\eta\xi\mu\nu,\alpha\beta\gamma\delta} = \frac{\partial E_{\eta\xi\mu\nu}}{\partial F_{\alpha\beta\gamma\delta}} = \delta_{\alpha\eta}\delta_{\beta\xi}\delta_{\gamma\mu}\delta_{\delta\nu}$$
$$-\frac{1}{7}\left(\delta_{\alpha\eta}\delta_{\beta\xi}\delta_{\gamma\delta}\delta_{\mu\nu} + \delta_{\alpha\eta}\delta_{\beta\delta}\delta_{\gamma\mu}\delta_{\xi\nu} + \delta_{\alpha\eta}\delta_{\beta\gamma}\delta_{\delta\nu}\delta_{\xi\mu}\right.$$
$$\left.+\delta_{\alpha\delta}\delta_{\beta\xi}\delta_{\gamma\mu}\delta_{\eta\nu} + \delta_{\alpha\gamma}\delta_{\beta\xi}\delta_{\delta\nu}\delta_{\eta\mu} + \delta_{\alpha\beta}\delta_{\gamma\mu}\delta_{\delta\nu}\delta_{\eta\xi}\right)$$
$$+\frac{1}{35}\left(\delta_{\alpha\beta}\delta_{\gamma\delta}\delta_{\eta\xi}\delta_{\mu\nu} + \delta_{\alpha\gamma}\delta_{\beta\delta}\delta_{\eta\mu}\delta_{\xi\nu} + \delta_{\alpha\delta}\delta_{\beta\gamma}\delta_{\eta\nu}\delta_{\xi\mu}\right)$$
(21c)

Similar to Eq. (22a) for transforming the traced $Q_{\alpha\beta}$ to the traceless $\Theta_{\alpha\beta}$, the traced octopole $O_{\alpha\beta\gamma}$ and hexadecapole $H_{\alpha\beta\gamma\delta}$ in Eqs. (5c–5d) can be transformed to the traceless octopole $\Omega_{\alpha\beta\gamma}$ and hexadecapole $\Phi_{\alpha\beta\gamma\delta}$ with the aid of Eqs. (21b–21c), i.e.,

$$\Omega_{\alpha\beta\gamma} = -f_{\Omega}6\frac{\partial U}{\partial F_{\alpha\beta\gamma}} = -\frac{15}{6}6\frac{\partial U}{\partial E_{\delta\eta\xi}}\frac{\partial E_{\delta\eta\xi}}{\partial F_{\alpha\beta\gamma}}$$
$$= \frac{1}{6}\left[15 O_{\alpha\beta\gamma} - 3\left(O_{\alpha\mu\mu}\delta_{\beta\gamma} + O_{\mu\beta\mu}\delta_{\alpha\gamma} + O_{\mu\mu\gamma}\delta_{\alpha\beta}\right)\right]$$
(22b)

$$\Phi_{\alpha\beta\gamma\delta} = -f_{\Phi}24\frac{\partial U}{\partial F_{\alpha\beta\gamma\delta}} = -\frac{105}{24}24\frac{\partial U}{\partial E_{\eta\xi\mu\nu}}\frac{\partial E_{\eta\xi\mu\nu}}{\partial F_{\alpha\beta\gamma\delta}}$$
$$= \frac{1}{24}\left[105 H_{\alpha\beta\gamma\delta} - 15\left(H_{\alpha\beta\mu\mu}\delta_{\gamma\delta} + H_{\alpha\mu\gamma\mu}\delta_{\beta\delta} + H_{\alpha\mu\mu\delta}\delta_{\beta\gamma} + H_{\mu\beta\gamma\mu}\delta_{\alpha\delta} + H_{\mu\beta\mu\delta}\delta_{\alpha\gamma} + H_{\mu\mu\gamma\delta}\delta_{\alpha\beta}\right)\right.$$
$$\left.+3\left(H_{\mu\mu\nu\nu}\delta_{\alpha\beta}\delta_{\gamma\delta} + H_{\mu\nu\mu\nu}\delta_{\alpha\gamma}\delta_{\beta\delta} + H_{\mu\nu\nu\mu}\delta_{\alpha\delta}\delta_{\beta\gamma}\right)\right]$$
(22c)

In Eqs. (22a–22c), the Buckingham factor of the transformation from traced $2^M$-moment to the traceless $2^M$-moment is

$$f_M = \frac{(2M-1)!!}{M!}$$
(23)

in which ! denotes factorial and !! denotes double factorial with $(2M-1)!! = (2M-1)\cdot(2M-3)\cdots 3\cdot 1$. Thus, $f_{\Omega} = \frac{15}{6}$ and $f_{\Phi} = \frac{105}{24}$ are used in

Eqs. (22b) and (22c), respectively. If $\rho(r)$ is known, the traceless multipoles can be written explicitly using Eqs. (3c–3e) and (22a–22c), i.e.,

$$\Theta_{\alpha\beta} = \frac{1}{2}\int_V dv \rho(r)\left(3r_\alpha r_\beta - r^2 \delta_{\alpha\beta}\right) \tag{24a}$$

$$\Omega_{\alpha\beta\gamma} = \frac{1}{6}\int_V dv \rho(r)\left[15 r_\alpha r_\beta r_\gamma - 3r^2\left(r_\alpha \delta_{\beta\gamma} + r_\beta \delta_{\alpha\gamma} + r_\gamma \delta_{\alpha\beta}\right)\right] \tag{24b}$$

$$\begin{aligned}\Phi_{\alpha\beta\gamma\delta} = \frac{1}{24}\int_V dv \rho(r)\Big[&105 r_\alpha r_\beta r_\gamma r_\delta - 15r^2\left(r_\alpha r_\beta \delta_{\gamma\delta} + r_\alpha r_\gamma \delta_{\beta\delta} + r_\alpha r_\delta \delta_{\beta\gamma}\right.\\&\left.+ r_\beta r_\gamma \delta_{\alpha\delta} + r_\beta r_\delta \delta_{\alpha\gamma} + r_\gamma r_\delta \delta_{\alpha\beta}\right) + 3r^4\left(\delta_{\alpha\beta}\delta_{\gamma\delta} + \delta_{\alpha\gamma}\delta_{\beta\delta} + \delta_{\alpha\delta}\delta_{\beta\gamma}\right)\Big]\end{aligned} \tag{24c}$$

Using the same reasoning as in Eq. (17), it can be verified by using either Eq. (22) or Eq. (24) for the following traceless identities

$$\Theta_{\alpha\alpha} = 0 \tag{25a}$$

$$\Omega_{\alpha\alpha\gamma} = \Omega_{\alpha\beta\alpha} = \Omega_{\alpha\beta\beta} = 0 \tag{25b}$$

$$\Phi_{\alpha\alpha\gamma\delta} = \Phi_{\alpha\beta\alpha\delta} = \Phi_{\alpha\beta\gamma\alpha} = \Phi_{\alpha\beta\beta\delta} = \Phi_{\alpha\beta\gamma\beta} = \Phi_{\alpha\beta\gamma\gamma} = 0 \tag{25c}$$

Though the transformation from the traced to the traceless multipoles is somehow tedious, the expression of energy is straight-forward, as shown in Eq. (20) for the quadrupole. It can be seen that the energy would be the same by multiplying and then dividing $f_M$ in Eq. (23). Extending the Eq. (20), which is for the permanent quadrupole, to the corresponding inducible counterpart in Eq. (11b), we have

$$\begin{aligned}U - U^0 = -\frac{1}{f_\Theta}\frac{1}{2}\bigg(&f_\Theta Q_{\alpha\beta} + f_\Theta a_{\gamma,\alpha\beta}E_\gamma + \frac{1}{2}f_\Theta b_{\gamma\delta,\alpha\beta}E_\gamma E_\delta + \frac{1}{2}f_\Theta c^{(2)}_{\gamma,\alpha\beta,\delta\eta}E_\gamma E_{\delta\eta}\\&+\frac{1}{2}f_\Theta c_{\alpha\beta,\gamma\delta}E_{\gamma\delta} + \frac{1}{f_\Omega}\frac{1}{3!}f_\Theta f_\Omega g_{\alpha\beta,\gamma\delta\eta}E_{\gamma\delta\eta} + \cdots\bigg)E_{\alpha\beta}\\= -\frac{1}{3}\bigg(&\Theta_{\alpha\beta} + A_{\gamma,\alpha\beta}E_\gamma + \frac{1}{2}B_{\gamma\delta,\alpha\beta}E_\gamma E_\delta + \frac{1}{2}C^{(2)}_{\gamma,\alpha\beta,\delta\eta}E_\gamma E_{\delta\eta}\\&+\frac{1}{2}C_{\alpha\beta,\gamma\delta}E_{\gamma\delta} + \frac{1}{15}G_{\alpha\beta,\gamma\delta\eta}E_{\gamma\delta\eta} + \cdots\bigg)E_{\alpha\beta}\end{aligned} \tag{26a}$$

in which $f_\Theta = \dfrac{3}{2}$ ~~as,~~ and $f_\Omega = \dfrac{15}{6}$, according to Eq. (23), and $A_{\gamma,\alpha\beta}$, $B_{\gamma\delta,\alpha\beta}$, $C_{\alpha\beta,\gamma\delta}$, $C^{(2)}_{\gamma,\alpha\beta,\delta\eta}$, $G_{\alpha\beta,\gamma\delta\eta}$ denote the traceless multipole-multipole polarizabilities, with the corresponding lowercase symbols denote the traced counterparts. Thus, the way to convert the energy from the expression of the traced multipoles and (hyper)polarizabilities to the expression of the corresponding traceless counterpart is straight forward, i.e., ~~by multiplying $f_\mu$ if electric field $E_\alpha$ is involved~~, multiplying $f_\Theta$ if electric field gradient $E_{\alpha\beta}$ is involved, and multiplying $f_\Omega$ if electric field gradient $E_{\alpha\beta\gamma}$ is involved, *etc.*. Such manipulation converts Eq. (11b) for the traced multipoles and multipole-multipole polarizabilities to the traceless counterparts in Eq. (26a). It is notable that for the terms involving quadrupole-quadrupole polarizability $c_{\alpha\beta,\gamma\delta}$ and dipole-quadrupole-quadrupole polarizability $c^{(2)}_{\gamma,\alpha\beta,\delta\eta}$, which have inherently exchangeable group in the sub-indices, the corresponding $f_\Theta$ is multiplied only once. This is because the corresponding traceless $C_{\alpha\beta,\gamma\delta}$ and $\underline{C^{(2)}_{\gamma,\alpha\beta,\delta\eta}}$ are obtained by a single differentiation with respect to $F_{\alpha\beta}$, and both groups are converted to the traceless form, as shown in Eq. (28) below. Moreover, it should be emphasized that Eq. (26a) merely denote $-\dfrac{1}{2}Q'_{\alpha\beta}E_{\alpha\beta} = -\dfrac{1}{3}\Theta'_{\alpha\beta}E_{\alpha\beta}$ by extending Eq. (20) to include inducible quadrupoles contributed from various external electric fields and gradients, but can not give the explicit form of the multipole-multipole polarizabilities. For example, Eq. (26a) denotes $-\dfrac{1}{4}c_{\gamma\delta,\alpha\beta}E_{\alpha\beta} = -\dfrac{1}{6}C_{\gamma\delta,\alpha\beta}E_{\alpha\beta}$, but can not give the explicit form of $C_{\gamma\delta,\alpha\beta}$, which will be derived below in Eq. (28).

Based on the above discussions regarding Eq. (26a), the contribution of the energy from the traceless octopole and hexadecapole, with inducible components, can be extended from Eqs. (11c–11d), i.e.,

$$U - U^0 = -\frac{1}{f_\Omega} \frac{1}{6} \left( f_\Omega O_{\alpha\beta\gamma} + f_\Omega d_{\delta,\alpha\beta\gamma} E_\delta + \frac{1}{f_\Theta} \frac{1}{2} f_\Theta f_\Omega g_{\delta\eta,\alpha\beta\gamma} E_{\delta\eta} + \frac{1}{3!} f_\Omega p_{\alpha\beta\gamma,\delta\eta\xi} E_{\delta\eta\xi} + \cdots \right) E_{\alpha\beta\gamma}$$

$$= -\frac{1}{15} \left( \Omega_{\alpha\beta\gamma} + D_{\delta,\alpha\beta\gamma} E_\delta + \frac{1}{3} G_{\delta\eta,\alpha\beta\gamma} E_{\delta\eta} + \frac{1}{6} P_{\alpha\beta\gamma,\delta\eta\xi} E_{\delta\eta\xi} + \cdots \right) E_{\alpha\beta\gamma}$$

(26b)

$$U - U^0 = -\frac{1}{f_\Phi} \frac{1}{24} \left( f_\Phi H_{\alpha\beta\gamma\delta} + \cdots \right) E_{\alpha\beta\gamma\delta} = -\frac{1}{105} \left( \Phi_{\alpha\beta\gamma\delta} + \cdots \right) E_{\alpha\beta\gamma\delta} \quad (26c)$$

in which $D_{\delta,\alpha\beta\gamma}$ and $P_{\delta\eta\xi,\alpha\beta\gamma}$ are the dipole-octopole and octopole-octopole polarizabilities, respectively, in the traceless form. Note that since the groups of the subscripts of octopole-octopole polarizabilities are inherently exchangeable, the term involving it is multiplied by $f_\Omega$ only once. Summing Eqs. (26a–26c), and skipping the same terms to avoid double counts, Buckingham expansion of the electrostatic energy of a polarizable charge distribution in the traced form, Eq. (12), can then be transform to the traceless form, i.e.,

$$\begin{aligned} U = U^0 &+ q\varphi \\ &- \left( \mu_\alpha E_\alpha + \frac{1}{2} \alpha_{\alpha\beta} E_\alpha E_\beta + \frac{1}{6} \beta_{\alpha\beta\gamma} E_\alpha E_\beta E_\gamma + \frac{1}{24} \gamma_{\alpha\beta\gamma\delta} E_\alpha E_\beta E_\gamma E_\delta + \cdots \right) \\ &- \frac{1}{3} \left[ \left( \Theta_{\alpha\beta} + A_{\gamma,\alpha\beta} E_\gamma + \frac{1}{2} B_{\gamma\delta,\alpha\beta} E_\gamma E_\delta + \cdots \right) + \frac{1}{2} \left( C_{\alpha\beta,\delta\eta} + C^{(2)}_{\gamma,\alpha\beta,\delta\eta} E_\gamma + \cdots \right) E_{\delta\eta} + \cdots \right] E_{\alpha\beta} \\ &- \frac{1}{15} \left( \Omega_{\alpha\beta\gamma} + D_{\delta,\alpha\beta\gamma} E_\delta + \frac{1}{3} G_{\delta\eta,\alpha\beta\gamma} E_{\delta\eta} + \frac{1}{6} P_{\delta\eta\xi,\alpha\beta\gamma} E_{\delta\eta\xi} + \cdots \right) E_{\alpha\beta\gamma} \\ &- \frac{1}{105} \Phi_{\alpha\beta\gamma\delta} E_{\alpha\beta\gamma\delta} - \cdots \end{aligned}$$

(27)

Eq. (27) is the final expression of the Buckingham expansion in the traceless form[3-5], in which all the multipoles and multipole-multipole polarizabilities are traceless.

The transformation from the traced multipoles to the traceless multipoles are given by Eqs. (22a–22c). We next derive the transformation from the traced multipole-multipole polarizabilities in Eqs. (6) and (8) to the traceless counterparts, using the traceless multipoles and the auxiliary traced electric field gradients. Using Eq. (27), we have

$$A_{\gamma,\alpha\beta} = -3\frac{\partial^2 U}{\partial F_{\alpha\beta}\partial E_\gamma} = \frac{\partial \Theta_{\alpha\beta}}{\partial E_\gamma} = \frac{1}{2}\left(3\frac{\partial Q_{\alpha\beta}}{\partial E_\gamma} - \frac{\partial Q_{\mu\mu}}{\partial E_\gamma}\delta_{\alpha\beta}\right) = \frac{1}{2}\left(3a_{\gamma,\alpha\beta} - a_{\gamma,\mu\mu}\delta_{\alpha\beta}\right) \quad (28a)$$

$$B_{\gamma\delta,\alpha\beta} = -3\frac{\partial^3 U}{\partial F_{\alpha\beta}\partial E_\gamma \partial E_\delta} = \frac{\partial A_{\gamma,\alpha\beta}}{\partial E_\delta} = \frac{1}{2}\left(3b_{\gamma\delta,\alpha\beta} - b_{\gamma\delta,\mu\mu}\delta_{\alpha\beta}\right) \quad (28b)$$

$$C_{\alpha\beta,\gamma\delta} = -3\frac{\partial^2 U}{\partial F_{\alpha\beta}\partial F_{\gamma\delta}} = \frac{\partial \Theta_{\alpha\beta}}{\partial F_{\gamma\delta}} = \frac{\partial \Theta_{\alpha\beta}}{\partial E_{\kappa\lambda}}\frac{\partial E_{\kappa\lambda}}{\partial F_{\gamma\delta}} = \frac{1}{2}\left(3\frac{\partial Q_{\alpha\beta}}{\partial E_{\kappa\lambda}} - \frac{\partial Q_{\mu\mu}}{\partial E_{\kappa\lambda}}\delta_{\alpha\beta}\right)J_{\kappa\lambda,\gamma\delta}$$
$$= \frac{1}{6}\left(9c_{\alpha\beta,\gamma\delta} - 3c_{\alpha\beta,\mu\mu}\delta_{\gamma\delta} - 3c_{\mu\mu,\gamma\delta}\delta_{\alpha\beta} + c_{\mu\mu,\nu\nu}\delta_{\alpha\beta}\delta_{\gamma\delta}\right) \quad (28c)$$

$$C^{(2)}_{\gamma,\alpha\beta,\delta\eta} = -3\frac{\partial^3 U}{\partial F_{\alpha\beta}\partial F_{\delta\eta}\partial E_\gamma} = \frac{\partial C_{\alpha\beta,\delta\eta}}{\partial E_\gamma}$$
$$= \frac{1}{6}\left(9c^{(2)}_{\gamma,\alpha\beta,\delta\eta} - 3c^{(2)}_{\gamma,\alpha\beta,\mu\mu}\delta_{\delta\eta} - 3c^{(2)}_{\gamma,\mu\mu,\delta\eta}\delta_{\alpha\beta} + c^{(2)}_{\gamma,\mu\mu,\nu\nu}\delta_{\alpha\beta}\delta_{\delta\eta}\right) \quad (28d)$$

$$D_{\delta,\alpha\beta\gamma} = -15\frac{\partial^2 U}{\partial F_{\alpha\beta\gamma}\partial E_\delta} = \frac{\partial \Omega_{\alpha\beta\gamma}}{\partial E_\delta}$$
$$= \frac{1}{6}\left[15d_{\delta,\alpha\beta\gamma} - 3\left(d_{\delta,\alpha\mu\mu}\delta_{\beta\gamma} + d_{\delta,\mu\beta\mu}\delta_{\alpha\gamma} + d_{\delta,\mu\mu\gamma}\delta_{\alpha\beta}\right)\right] \quad (28e)$$

$$G_{\delta\eta,\alpha\beta\gamma} = -45\frac{\partial^2 U}{\partial F_{\alpha\beta\gamma}\partial F_{\delta\eta}} = 3\frac{\partial \Omega_{\alpha\beta\gamma}}{\partial F_{\delta\eta}} = 3\frac{\partial \Omega_{\alpha\beta\gamma}}{\partial E_{\kappa\lambda}}J_{\kappa\lambda,\delta\eta}$$
$$= \frac{1}{4}\left[15g_{\kappa\lambda,\alpha\beta\gamma} - 3\left(g_{\kappa\lambda,\alpha\mu\mu}\delta_{\beta\gamma} + g_{\kappa\lambda,\mu\beta\mu}\delta_{\alpha\gamma} + g_{\kappa\lambda,\mu\mu\gamma}\delta_{\alpha\beta}\right)\right]\left(\delta_{\delta\kappa}\delta_{\eta\lambda} - \frac{1}{3}\delta_{\delta\eta}\delta_{\kappa\lambda}\right) \quad (28f)$$
$$= \frac{1}{4}\left[15g_{\delta\eta,\alpha\beta\gamma} - 5g_{\mu\mu,\alpha\beta\gamma}\delta_{\delta\eta} - 3\left(g_{\delta\eta,\alpha\nu\nu}\delta_{\beta\gamma} + g_{\delta\eta,\nu\beta\nu}\delta_{\alpha\gamma} + g_{\delta\eta,\nu\nu\gamma}\delta_{\alpha\beta}\right)\right.$$
$$\left.+ g_{\mu\mu,\alpha\nu\nu}\delta_{\beta\gamma}\delta_{\delta\eta} + g_{\mu\mu,\nu\beta\nu}\delta_{\alpha\gamma}\delta_{\delta\eta} + g_{\mu\mu,\nu\nu\gamma}\delta_{\alpha\beta}\delta_{\delta\eta}\right]$$

$$P_{\alpha\beta\gamma,\delta\eta\xi} = -45 \frac{\partial^2 U}{\partial F_{\alpha\beta\gamma} \partial F_{\delta\eta\xi}} = 3 \frac{\partial \Omega_{\alpha\beta\gamma}}{\partial F_{\delta\eta\xi}} = 3 \frac{\partial \Omega_{\alpha\beta\gamma}}{\partial E_{\lambda\mu\nu}} J_{\lambda\mu\nu,\delta\eta\xi}$$

$$= \frac{1}{6}\Big[15 p_{\alpha\beta\gamma,\delta\eta\xi}$$

$$-3\big(p_{\alpha\beta\gamma,\delta\mu\mu}\delta_{\eta\xi} + p_{\alpha\beta\gamma,\mu\eta\mu}\delta_{\delta\xi} + p_{\alpha\beta\gamma,\mu\mu\xi}\delta_{\delta\eta} + p_{\alpha\mu\mu,\delta\eta\xi}\delta_{\beta\gamma} + p_{\mu\beta\mu,\delta\eta\xi}\delta_{\alpha\gamma} + p_{\mu\mu\gamma,\delta\eta\xi}\delta_{\alpha\beta}\big)$$

$$+\frac{1}{5}\big(p_{\alpha\mu\mu,\delta\nu\nu}\delta_{\beta\gamma}\delta_{\eta\xi} + p_{\mu\beta\mu,\delta\nu\nu}\delta_{\alpha\gamma}\delta_{\eta\xi} + p_{\mu\mu\gamma,\delta\nu\nu}\delta_{\alpha\beta}\delta_{\eta\xi} + p_{\alpha\mu\mu,\nu\eta\nu}\delta_{\beta\gamma}\delta_{\delta\xi} + p_{\mu\beta\mu,\nu\eta\nu}\delta_{\alpha\gamma}\delta_{\delta\xi}$$

$$+p_{\mu\mu\gamma,\nu\eta\nu}\delta_{\alpha\beta}\delta_{\delta\xi} + p_{\alpha\mu\mu,\nu\nu\xi}\delta_{\beta\gamma}\delta_{\delta\eta} + p_{\mu\beta\mu,\nu\nu\xi}\delta_{\alpha\gamma}\delta_{\delta\eta} + p_{\mu\mu\gamma,\nu\nu\xi}\delta_{\alpha\beta}\delta_{\delta\eta}\big)\Big]$$

(28g)

in which chain rule is used in order to get the traceless multipole-multipole polarizabilities, with the aid of Eqs. (21a–21b). Note that some of the transformations in Eq. (28), for example, Eq. (28c), have also been given by Applequist[13] using a different approach. It can be verified using Eq. (28) the following traceless identities

$$A_{\gamma,\alpha\alpha} = 0 \tag{29a}$$

$$B_{\gamma\delta,\alpha\alpha} = 0 \tag{29b}$$

$$C_{\alpha\alpha,\gamma\delta} = C_{\alpha\beta,\gamma\gamma} = 0 \tag{29c}$$

$$C^{(2)}_{\gamma,\alpha\alpha,\delta\eta} = C^{(2)}_{\gamma,\alpha\beta,\delta\delta} = 0 \tag{29d}$$

$$D_{\delta,\alpha\alpha\gamma} = D_{\delta,\alpha\beta\alpha} = D_{\delta,\alpha\beta\beta} = 0 \tag{29e}$$

$$G_{\delta\eta,\alpha\alpha\gamma} = G_{\delta\eta,\alpha\beta\alpha} = G_{\delta\eta,\alpha\beta\beta} = G_{\delta\delta,\alpha\beta\gamma} = 0 \tag{29f}$$

$$P_{\alpha\alpha\gamma,\delta\eta\xi} = P_{\alpha\beta\alpha,\delta\eta\xi} = P_{\alpha\beta\beta,\delta\eta\xi} = P_{\alpha\beta\gamma,\delta\delta\xi} = P_{\alpha\beta\gamma,\delta\eta\delta} = P_{\alpha\beta\gamma,\delta\eta\eta} = 0 \tag{29g}$$

In getting the above identities, using the first identity $D_{\delta,\alpha\alpha\gamma}$ in Eq. (29e) as example, we have from Eq. (28e)

$$D_{\delta,\alpha\alpha\gamma} = \frac{1}{6}\Big[15 d_{\delta,\alpha\alpha\gamma} - 3\big(d_{\delta,\alpha\mu\mu}\delta_{\alpha\gamma} + d_{\delta,\mu\alpha\mu}\delta_{\alpha\gamma} + d_{\delta,\mu\mu\gamma}\delta_{\alpha\alpha}\big)\Big]$$

$$= \frac{1}{6}\Big[15 d_{\delta,\mu\mu\gamma} - 3\big(d_{\delta,\gamma\mu\mu} + d_{\delta,\mu\gamma\mu} + 3 d_{\delta,\mu\mu\gamma}\big)\Big] = 0$$

in which we use the fact that the repeated sub-index $\alpha$ is dummy, so that $\delta_{\alpha\alpha} = 3$ and $\alpha$ can be replaced by another dummy variable $\mu$, i.e., $D_{\delta,\alpha\alpha\gamma} = D_{\delta,\mu\mu\gamma}$; and $D_{\delta,\gamma\mu\mu} = D_{\delta,\mu\gamma\mu} = D_{\delta,\mu\mu\gamma}$ since the multipole-multipole polarizabilities are invariant upon the permutation of the sub-indices in the same group. Similar arguments apply on the identities in Eqs. (29e–29g). From the above discussions, the traceless feature of the multipole-multipole polarizabilities in Eq. (28) is the consequence of the traceless electric field gradients, attributed to the Laplace equation.

## 4. Numerical Multipoles and (Hyper)polarizabilities with Finite Field on *Ab Initio* Calculations. A Case Study on Carbon Dioxide.

Carbon dioxide has attracted much interest, not only for its importance in industry and environmental science, but also for its unique electronic properties as a quadrupole molecule. Buckingham *et. al.*[3, 7] had designed a four-wires condenser to measure the accurate quadrupole of $CO_2$, which laid the foundation for the electric-field-gradient-induced birefringence (EFGIB) experiments[17-19]. The mean and anisotropy dipole polarizability of $CO_2$ could be measured by the depolarization ratio of Rayleigh scattering[20]. Ritchie *et. al.*[21] utilized the Kerr effect to measure the mean second dipole hyperpolarizability, suggested by Buckingham and Pople[6]. On the other hand, higher order multipoles and (hyper)polarizabilities are difficult to directly measure from the experiments nowadays.

Computationally, Maroulis and co-workers have calculated the multipoles and (hyper)polarizabilities of $CO_2$ numerically against high-level *ab initio* calculations with both charge perturbation method (CPM)[22] and finite field method (FFM)[10, 23]. In the charge perturbation method, charge is placed at certain distance $R$ from the molecule. Since all orders of electric field and gradients are generated by the charge, according to Eq. (14), the finite difference is based on the full Buckingham expansion. Moreover, for a polyatom molecule, the point multipole representations are valid only at $R\rightarrow\infty$. Thus, in the actual calculations, the multipoles and (hyper)polarizabilities are often calculated at several different $R$'s, and then extrapolated to $R\rightarrow\infty$ value to refine the result[8]. Such difficulty of CPM can be overcome by FFM, in which the electric field or gradient tensors are manipulated directly to generate the desired uniform external field or gradients. In this study, we apply FFM to calculate the multipoles and (hyper)polarizabilities of $CO_2$ numerically, as detailed in Appendix B.

The equilibrium geometry of an electronic ground state $CO_2$, of $D_{\infty h}$ point group, has only one structural degree of freedom, i.e., the C–O bond length $r_{CO}$. In this study, we take $r_{CO}$ = 2.196 $a_0$ (bohr), as deduced from precise experimental rotational constant of $CO_2$[24]. The cartesian coordinate system is set with the z-axis along the $D_{\infty h}$ symmetric axis and x, y, z-axes mutually perpendicular, corresponding to the principal axes of both $\alpha_{\alpha\beta}$ and $\Theta_{\alpha\beta}$. In the principal frame, there are 2 independent components of $\alpha_{\alpha\beta}$, 1 independent component of $\Theta_{\alpha\beta}$, and 3 independent components of $C_{\alpha\beta,\gamma\delta}$, respectively[5]. With the above geometry and coordinate, the *ab initio* calculations, with finite external electric field or gradient applied, were performed at CCSD(T)/daug-cc-pvqz level of theory using Gaussian16 package[15].

According to Eq. (6a), with $U$ in Eq. (12) or (27), the dipole polarizability can be calculate numerically with finite difference, i.e.[25],

$$\alpha_{\alpha\beta} = \frac{4}{3} \times \frac{U(-E_\alpha, E_\beta) + U(E_\alpha, -E_\beta) - U(E_\alpha, E_\beta) - U(-E_\alpha, -E_\beta)}{4 E_\alpha E_\beta} \\ -\frac{1}{3} \times \frac{U(-2E_\alpha, 2E_\beta) + U(2E_\alpha, -2E_\beta) - U(2E_\alpha, 2E_\beta) - U(-2E_\alpha, -2E_\beta)}{16 E_\alpha E_\beta} + O(E_\alpha^4) \quad (30)$$

in which $U(E_\alpha, E_\beta)$ denotes the electronic energy calculated at CCSD(T)/daug-cc-pvqz level of theory with finite field of $E_\alpha = E_\beta = \pm 0.001\ e/a_0^2$ applied on $\alpha$ and $\beta$ directions simultaneously. Eq. (30) is derived via Taylor series with solely electric fields applied, as a special case of the Buckingham expansion. Since $CO_2$ is orientated along the principal axes of polarizability with z-axis along the $D_{\infty h}$ symmetric axis, the calculations on $\alpha_{zz}$ and $\alpha_{xx}$ are simpler for the condition $\alpha = \beta$. In this case, $U(-E_\alpha, E_\alpha) = U_0$ corresponds to the energy without field, and $U(E_\alpha, E_\alpha) = U(2E_\alpha)$, *etc.*, so that 5 *ab initio* calculations are enough, instead of 8 *ab initio* calculations, in order to calculate each of $\alpha_{zz}$ and $\alpha_{xx}$ with FFM.

Similarly, the traced quadrupole and quadrupole-quadrupole polarizability can be calculated with

$$Q_{\alpha\beta} = \frac{4}{3} \times \frac{U(-E_{\alpha\beta}) - U(E_{\alpha\beta})}{2E_{\alpha\beta}} - \frac{1}{3} \times \frac{U(-2E_{\alpha\beta}) - U(2E_{\alpha\beta})}{4E_{\alpha\beta}} + O(E_{\alpha\beta}^4) \quad (31)$$

$$c_{\alpha\beta,\gamma\delta} = \frac{4}{3} \times \frac{U(-E_{\alpha\beta}, E_{\gamma\delta}) + U(E_{\alpha\beta}, -E_{\gamma\delta}) - U(E_{\alpha\beta}, E_{\gamma\delta}) - U(-E_{\alpha\beta}, -E_{\gamma\delta})}{8E_{\alpha\beta}E_{\gamma\delta}}$$
$$-\frac{1}{3} \times \frac{U(-2E_{\alpha\beta}, 2E_{\gamma\delta}) + U(2E_{\alpha\beta}, -2E_{\gamma\delta}) - U(2E_{\alpha\beta}, 2E_{\gamma\delta}) - U(-2E_{\alpha\beta}, -2E_{\gamma\delta})}{32E_{\alpha\beta}E_{\gamma\delta}} + O(E_{\alpha\beta}^4)$$

$$(32)$$

Detailed derivation of the above two finite field expressions are given by Eqs. (B5) and (B8) in Appendix B. The applied finite field gradient is $E_{\alpha\beta} = 0.0005\ e/a_0^3$. Since the symmetric $E_{\alpha\beta}$ tensor is manipulated in Gaussian16 package[15] with $E_{\beta\alpha} = E_{\alpha\beta} = 0.0005\ e/a_0^3$ for $\alpha \neq \beta$ and $E_{\alpha\alpha} = 0.001\ e/a_0^3$ for $\alpha = \beta$, the actual calculation with FFM gives $Q_{\alpha\beta} + Q_{\beta\alpha} = 2Q_{\alpha\beta}$. Thus, the result needs to be further divided by 2, as comparing Eq. (31) with Eq. (B5). For the same reason, the actual calculation with FFM gives $c_{\alpha\beta,\gamma\delta} + c_{\beta\alpha,\gamma\delta} + c_{\alpha\beta,\delta\gamma} + c_{\beta\alpha,\delta\gamma} = 4c_{\alpha\beta,\gamma\delta}$. Thus, the result needs to be further divided by 4, as comparing Eq. (32) with Eq. (B8). The corresponding traceless quadrupole $\Theta_{\alpha\beta}$ and quadrupole-quadrupole polarizability $C_{\alpha\beta,\gamma\delta}$ can be obtained via Eqs. (22a) and (28c), respectively.

**Table 1**. The dipole polarizability of $CO_2$ in atomic unit ($a_0^3$).

|  | CCSD(T)[a] | MP4SDQ[b] | CCSD(T)[c] | CCSD(T)[d] | Experiments | |
|---|---|---|---|---|---|---|
| $\alpha_{zz}$ | 26.740 | 27.140 | 27.07 | 26.8184 | 27.2408[e], 27.2909[g] | 27.3056[f], |
| $\alpha_{xx}$[h] | 12.852 | 12.869 | 12.99 | 12.8966 | 13.0018[e], 13.0379[g] | 13.0162[f], |
| $\bar{\alpha}$[i] | 17.481 | 17.626 | 17.69 | 17.5372 | 17.7481[e], 17.7889[g] | 17.7793[f], |
| $\Delta\alpha$[j] | 13.887 | 14.271 | 14.08 | 13.9217 | 14.2390[e], 14.2530[g] | 14.2894[f], |

[a]This work, FFM against CCSD(T)/daug-cc-pvqz with 315 contracted gaussian-type

basis functions (CGTBFs); [b]Ref [22], CPM against MP4SDQ/[6s4p3d1f] with 120 CGTBFs; [c]Ref [23], FFM against CCSD(T)/[6s4p4d1f] with 135 CGTBFs; [d]Ref [10], FFM against CCSD(T)/[7s5p4d2f] with 168 CGTBFs; [e]Ref [26], derived from the depolarization ratio of Rayleigh scattering; [f]Ref[20], derived from the depolarization ratio of Rayleigh scattering; [g]Ref [27], derived from the depolarization ratio of Rayleigh scattering; [h]For $CO_2$ orientated with z-axis along the $D_{\infty h}$ symmetry axis, $\alpha_{yy} = \alpha_{xx}$; [i]Isotropic (average) dipole polarizability, $\bar{\alpha} = (\alpha_{zz} + 2\alpha_{xx})/3$; [j]Anisotropic dipole polarizability, $\Delta\alpha = \alpha_{zz} - \alpha_{xx}$.

**Table 2**. The traceless quadrupole of $CO_2$ in atomic unit ($e \cdot a_0^2$).

| | CCSD(T)[a] | MP4SDQ[b] | CCSD(T)[c] | CCSD(T)[d] | Experiments |
|---|---|---|---|---|---|
| $\Theta_{zz}$[i] | -3.163 | -3.239 | -3.19 | -3.17 | -3.0482[e], -3.1873[f], -3.1806[g], -3.1895[h] |

[a]This work; [b]Ref [22]; [c]Ref [23]; [d]Ref [28], FFM against CCSD(T)/[8s5p5d2f] with 186 CGTBFs; [e]Ref [7], EFGIB experiment proposed by Buckingham; [f]Ref [18], EFGIB experiment; [g]Ref[19], EFGIB experiment; [h]Ref[17], EFGIB experiment; [i]For $CO_2$ orientated with z-axis along the $D_{\infty h}$ symmetry axis, the traceless condition in Eq. (25a) implies that $\Theta_{xx} = \Theta_{yy} = -\Theta_{zz}/2$.

**Table 3**. The traceless quadrupole-quadrupole polarizability of $CO_2$ in atomic unit ($a_0^5$).

| | CCSD(T)[a] | MP4SDQ[b] | CCSD(T)[c] | CCSD(T)[d] |
|---|---|---|---|---|
| $C_{zz,zz}$[e] | 80.878 | 81.079 | 80.94 | 81.14 |
| $C_{xx,xx}$[f] | 34.082 | 33.021 | 34.13 | 33.97 |
| $C_{xz,xz}$[g] | 53.506 | 54.039 | 54.81 | 53.88 |
| $\bar{C}$[h] | 78.158 | 77.757 | 79.25 | 78.39 |

[a]This work; [b]Ref [22]; [c]Ref [23]; [d]Ref [10]; [e]For $CO_2$ of $D_{\infty h}$ symmetry, the traceless condition in in Eq. (29c) implies that $C_{zz,xx} = C_{zz,yy} = -C_{zz,zz}/2$. Also, as discussed in Sec. 2, the symmetry of subscripts implies that $C_{xx,zz} = C_{zz,xx} = C_{yy,zz} = C_{zz,yy}$; [f]$C_{xx,yy} = C_{yy,xx} = -C_{xx,xx} - C_{xx,zz}$; [g]$C_{yz,yz} = C_{xz,xz}$, $C_{xy,xy} = C_{xx,xx} - C_{zz,zz}/4$, and the values are invariant under the permutations of subscripts yz or xz or xy within the same group. All the other non-mentioned $C_{\alpha\beta,\gamma\delta}$'s are identically zeros; [h]Isotropic quadrupole-quadrupole polarizability $\bar{C} = (C_{zz,zz} + 8C_{xx,xx} + 8C_{xz,xz})/10$[22].

Tables 1, 2, and 3 list the independent components of dipole polarizability, quadrupole, and quadrupole-quadrupole polarizability, respectively. Comparisons with high level *ab initio* calculations of both CPM and FFM, as well as the experimental measurements are also listed for dipole polarizability and quadrupole in

Tables 1 and 2. It can be seen that the agreement of current study with previous computational and experimental studies is reasonable. Thus, the general Buckingham expansion in the traced form in Eq. (B1) is validated. The goodness of Eq. (B1) is that it can be applied to calculate all the multipoles and (hyper)polarizabilities with accuracy of $O(E^n)$ for arbitrary order $n$ in a consistent manner with FFM. Of course, the accuracy of FFM is also highly dependent on the level of theory and basis set in the *ab initio* calculations, in order to get an accurate $U(E)$. For further study, we plan to adopt high level *ab initio* calculations and extrapolate to the complete basis set limit to get higher accuracy of $U(E)$.

## 5. Summary.

In this study, we derive Buckingham expansion[3-5] in both traced and traceless forms. Specifically, with explicitly taking the Buckingham convention, aiming on the final form of Buckingham expansion in its widely accepted form, we derive in Sec. 2 Buckingham expansion in the traced form, using Taylor series. Based on the above derivation, a general Buckingham expansion in the traced form, Eq. (B1), is proposed. Based on Eq. (B1), numerical calculations of the multipoles and (hyper)polarizabilities with finite field method, with accuracy of $O(E^4)$, in which $E$ denotes the externally applied finite fields or gradients, is derived. The transformation from the traced multipoles and multipole-multipole polarizabilities to the corresponding traceless counterparts can be performed with an auxiliary traced electric field gradient, as described in Sec. 3. The numerical calculations based on finite field method against high level *ab initio* calculations of $CO_2$, as a case study in Sec. 4, are in good agreements with experimental results and previous theoretical calculations. We hope the method in the current study can be adopted to in the calculations molecular/ionic multipoles and (hyper)polarizabilities, and further studies are in progress.

**Acknowledgments**. This work was supported by the National Natural Science Foundation of China (Grant Nos. 21573112, 21421001).

**Conflict of interest**. The authors declare that they have no conflict of interest.

**Appendix A. Comparison with previous works.**

Buckingham proposed the Buckingham expansion in the traceless form[3-5], but did not provide a complete derivation to the best of our knowledge. McLean and Yoshimine (M&Y)[12], in communications with Buckingham, derived Buckingham expansion in the traced form with multi-variable Taylor series, and then extended to the traceless form by solving the traceless multipoles using Laplace equation, and solving the traceless multipole-multipole polarizabilities using the traceless features in Eq. (29). Our derivation also starts from Taylor series, as presented in Sec. 2. On the other hand, the main difference is that the traceless multipoles and multipole-multipole polarizabilities are derived in Sec. 3 in this work. Thus, Eq. (29), as the consequence of Laplace equation, is not needed in deriving the traceless multipole-multipole polarizabilities. Since M&Y's derivation[12] was adopted in many studies in numerical calculations of molecular multipoles and (hyper)polarizabilities[23, 29], we give in this appendix mainly comparison with M&Y's derivation[12].

Firstly, we need to clarify the convention in M&Y's derivation[12]. For a charge distribution of $\rho(r)$ centered at $O$, represented by point multipoles, the energy $u$, which is denoted as $U$ in our study, is a function of electric potential, fields, and gradients, i.e.[12]

$$u = u\left(\varphi, E_x, E_y, E_z, E_{xx}, E_{xy}, E_{xz}, E_{yy}, E_{yz}, E_{zz}, \right.$$
$$\left. E_{xxx}, E_{xxy}, E_{xxz}, E_{xyy}, E_{xyz}, E_{xzz}, E_{yyy}, E_{yyz}, E_{yzz}, E_{zzz}, E_{xxxx}, \cdots\right) \quad (A1)$$

in which the energy $u_0$ in zero external field is set to be zero without losing generality. It is notable that only half of the off-diagonal electric field gradient tensor components are included in the above equation. For example, $E_{xy}$ is included in Eq. (A1), but $E_{yx}$ is not included, since field gradient is symmetric and $E_{xy} = E_{yx}$. Thus, the convention in M&Y's derivation[12] is that only one of the symmetric electric field gradient components is included in the energy expression. Similarly, by such convention, only one out of the three permutations of, the *xxy* components is include for $E_{xxy}$, and one out of the six permutations of the *xyz* components is included for $E_{xyz}$, *etc.*. Specifically, for a *l*th rank field gradient tensor, $E_{\alpha\beta\cdots}$, only one out of the $d_{\alpha\beta\cdots}$ permutations is included in Eq. (A1), with

$$d_{\alpha\beta\cdots} = \frac{l!}{l_x! l_y! l_z!} \quad (A2)$$

in which $l_x$, $l_y$, and $l_z$ are the number of the repeated subscripts $\{x, y, z\}$, respectively, and $l = l_x + l_y + l_z$. By such convention, it is understood that only the independent field gradient components, i.e., 6 out of 9 $E_{\alpha\beta}$, 10 out of 27 $E_{\alpha\beta\gamma}$, 15 out of 81 $E_{\alpha\beta\gamma\delta}$, *etc.*, are included in Eq. (A1). Generally, the number of independent components for a *l*th rank tensor is $\sum_{i=1}^{l+1} i$. Such convention is in distinct difference than the Buckingham convention, in which all the permutations of $E_{\alpha\beta\cdots}$ are included in the energy expression, as discussed in Sec. 2.

Nevertheless, following the convention of independent field gradient components of M&Y[12], Eq. (A1) is then expanded via Taylor series in the space of multiple variables, i.e.,

$$u = q\varphi + \sum_{\alpha} u'_{\alpha} E_{\alpha} + \sum_{\alpha} \sum_{\beta \geq \alpha} u'_{\alpha\beta} E_{\alpha\beta} + \sum_{\alpha} \sum_{\beta \geq \alpha} \sum_{\gamma \geq \beta} u'_{\alpha\beta\gamma} E_{\alpha\beta\gamma} + \sum_{\alpha} \sum_{\beta \geq \alpha} \sum_{\gamma \geq \beta} \sum_{\delta \geq \gamma} u'_{\alpha\beta\gamma\delta} E_{\alpha\beta\gamma\delta} + \cdots$$

$$+ \frac{1}{2} \sum_{\alpha} \sum_{\beta} u''_{\alpha\beta} E_{\alpha} E_{\beta} + \frac{1}{2} \sum_{\gamma} \sum_{\alpha} \sum_{\beta \geq \alpha} \left( u''_{\gamma,\alpha\beta} E_{\gamma} E_{\alpha\beta} + u''_{\alpha\beta,\gamma} E_{\alpha\beta} E_{\gamma} \right)$$

$$+ \frac{1}{2} \sum_{\alpha} \sum_{\beta \geq \alpha} \sum_{\gamma} \sum_{\delta \geq \gamma} u''_{\alpha\beta,\gamma\delta} E_{\alpha\beta} E_{\gamma\delta} + \frac{1}{2} \sum_{\delta} \sum_{\alpha} \sum_{\beta \geq \alpha} \sum_{\gamma \geq \beta} \left( u''_{\delta,\alpha\beta\gamma} E_{\delta} E_{\alpha\beta\gamma} + u''_{\alpha\beta\gamma,\delta} E_{\alpha\beta\gamma} E_{\delta} \right) + \cdots$$

$$+ \frac{1}{3!} \sum_{\alpha} \sum_{\beta} \sum_{\gamma} u'''_{\alpha\beta\gamma} E_{\alpha} E_{\beta} E_{\gamma} + \frac{1}{3!} \sum_{\alpha} \sum_{\beta \geq \alpha} \sum_{\gamma} \sum_{\delta} \left( u'''_{\alpha\beta,\gamma\delta} E_{\alpha\beta} E_{\gamma} E_{\delta} + u'''_{\gamma,\alpha\beta,\delta} E_{\gamma} E_{\alpha\beta} E_{\delta} + u'''_{\gamma\delta,\alpha\beta} E_{\gamma} E_{\delta} E_{\alpha\beta} \right) + \cdots$$

$$+ \frac{1}{4!} \sum_{\alpha} \sum_{\beta} \sum_{\gamma} \sum_{\delta} u''''_{\alpha\beta\gamma\delta} E_{\alpha} E_{\beta} E_{\gamma} E_{\delta} + \cdots$$

(A3)

in which the summation symbol $\sum$ is explicitly written, because M&Y's convention imposes the constraints on the range of subscripts for the $l$th rank field gradient tensors for $l \geq 2$. Apart from that, the exchange of the group of the subscripts, for example, in the constraint summation over $u''_{\gamma,\alpha\beta} E_{\alpha} E_{\alpha\beta} + u''_{\alpha\beta,\gamma} E_{\alpha\beta} E_{\gamma}$ in the second line of Eq. (A3), occurs as a consequence of the Taylor series of multi-variables. It is also notable that due to the inherent exchangeable groups of the sub-indices, the summations over $u''_{\gamma\delta,\alpha\beta} E_{\gamma\delta} E_{\alpha\beta}$, have already included in the summations over $u''_{\alpha\beta,\gamma\delta} E_{\alpha\beta} E_{\gamma\delta}$. Also, due to the free permutations of sub-indices $\gamma$ and $\delta$ in $u'''_{\alpha\beta,\gamma\delta}$, the summations over $u'''_{\alpha\beta,\delta\gamma} E_{\alpha\beta} E_{\delta} E_{\gamma}$, have already included in the summations over $u'''_{\alpha\beta,\gamma\delta} E_{\alpha\beta} E_{\gamma} E_{\delta}$, and similar for the other two terms involving $u'''_{\gamma,\alpha\beta,\delta}$ and $u'''_{\gamma\delta,\alpha\beta}$. Thus, all the permutations of the groups of subscripts are allowed, and such convention is also in distinct difference than the Buckingham convention discussed in Sec. 2, and a conversion to Buckingham convention is needed.

Eq. (A3) is the key to understand M&Y's derivation[12] to solve the multipoles and (hyper)polarizabilities. Briefly, the approach taken in M&Y's derivation[12] is that they firstly released the constraint on the independent electric field gradient variables in Eq. (A1), and then added a linear combinations of the traceless electric field gradients, i.e., Eq. (17), into Eq. (A3). This manipulation does not alter $u$, but Eq.

(A3) can be transformed to the Buckingham expansion in the traceless form, i.e., Eq. (27). Next, the traceless features of the multipole-multipole polarizabilities, i.e., Eq. (29), were invited to solve the coefficients in the linear combination. By equating the coefficients, the individual cartesian components of the multipoles and (hyper)polarizabilities were expressed in the derivatives of $u$ wrt electric fields or gradients[12].

Here, we take a simple route to get the same results in M&Y's derivation[12], by taking advantage of some of the derived expressions in Sec. 2 and Sec. 3. Since the multi-variable Taylor series in Eq. (A3) ought to give the same energy as the Buckingham expansion, by comparing equivalent terms of Eq. (A3) with the Buckingham expansion in the traced form in Eq. (12), and considering the symmetric permutations, we have

$$u'_\alpha = \frac{\partial u}{\partial E_\alpha} = -\mu_\alpha \tag{A4a}$$

$$u'_{\alpha\beta} = \frac{\partial u}{\partial E_{\alpha\beta}} = -\frac{d_{\alpha\beta}}{2} Q_{\alpha\beta} \tag{A4b}$$

$$u'_{\alpha\beta\gamma} = \frac{\partial u}{\partial E_{\alpha\beta\gamma}} = -\frac{d_{\alpha\beta\gamma}}{3!} O_{\alpha\beta\gamma} \tag{A4c}$$

$$u'_{\alpha\beta\gamma\delta} = \frac{\partial u}{\partial E_{\alpha\beta\gamma\delta}} = -\frac{d_{\alpha\beta\gamma\delta}}{4!} H_{\alpha\beta\gamma\delta} \tag{A4d}$$

$$u''_{\alpha\beta} = \frac{\partial^2 u}{\partial E_\alpha \partial E_\beta} = -\alpha_{\alpha\beta} \tag{A4e}$$

$$u''_{\gamma,\alpha\beta} = u''_{\alpha\beta,\gamma} = \frac{\partial^2 u}{\partial E_\gamma \partial E_{\alpha\beta}} = -\frac{d_{\alpha\beta}}{2} a_{\gamma,\alpha\beta} \tag{A4f}$$

$$u''_{\alpha\beta,\gamma\delta} = \frac{\partial^2 u}{\partial E_{\alpha\beta} \partial E_{\gamma\delta}} = -\frac{d_{\alpha\beta} d_{\gamma\delta}}{2} c_{\alpha\beta,\gamma\delta} \tag{A4g}$$

$$u''_{\delta,\alpha\beta\gamma} = u''_{\alpha\beta\gamma,\delta} = \frac{\partial^2 u}{\partial E_\delta \partial E_{\alpha\beta\gamma}} = -\frac{d_{\alpha\beta\gamma}}{6} d_{\delta,\alpha\beta\gamma} \tag{A4h}$$

$$u'''_{\alpha\beta\gamma} = \frac{\partial^3 u}{\partial E_\alpha \partial E_\beta \partial E_\gamma} = -\beta_{\alpha\beta\gamma} \tag{A4i}$$

$$u'''_{\gamma\delta,\alpha\beta} = u'''_{\gamma,\alpha\beta,\delta} = u'''_{\alpha\beta,\gamma\delta} = \frac{\partial^3 u}{\partial E_\gamma \partial E_\delta \partial E_{\alpha\beta}} = -\frac{d_{\alpha\beta}}{2} b_{\gamma\delta,\alpha\beta} \tag{A4j}$$

$$u''''_{\alpha\beta\gamma\delta} = \frac{\partial^4 u}{\partial E_\alpha \partial E_\beta \partial E_\gamma \partial E_\delta} = -\gamma_{\alpha\beta\gamma\delta} \tag{A4k}$$

It is notable that Eqs. (A4a), (A4e), (A4i), and (A4k) give the dipole, polarizability, and first and second polarizability, as in Eqs. (5a), (6a), (8a), and (8b). For the others, they give the relations between the derivatives of $u$ and the traced multipoles and multipole-multipole polarizabilities in Eqs. (5b–5d), (6b–6d), and (8c). We can next convert the traced multipoles and multipole-multipole polarizabilities to the traceless counterparts using Eqs. (22) and (28), and the results are

$$\Theta_{\alpha\beta} = -3\frac{u'_{\alpha\beta}}{d_{\alpha\beta}} + u'_{\mu\mu}\delta_{\alpha\beta} \tag{A5a}$$

$$\Omega_{\alpha\beta\gamma} = -15\frac{u'_{\alpha\beta\gamma}}{d_{\alpha\beta\gamma}} + 3\left(\frac{u'_{\alpha\mu\mu}}{d_{\alpha\mu\mu}}\delta_{\beta\gamma} + \frac{u'_{\mu\beta\mu}}{d_{\mu\beta\mu}}\delta_{\alpha\gamma} + \frac{u'_{\mu\mu\gamma}}{d_{\mu\mu\gamma}}\delta_{\alpha\beta}\right) \tag{A5b}$$

$$\Phi_{\alpha\beta\gamma\delta} = -105\frac{u'_{\alpha\beta\gamma\delta}}{d_{\alpha\beta\gamma\delta}} + 15\left(\frac{u'_{\alpha\beta\mu\mu}}{d_{\alpha\beta\mu\mu}}\delta_{\gamma\delta} + \frac{u'_{\alpha\mu\gamma\mu}}{d_{\alpha\mu\gamma\mu}}\delta_{\beta\delta} + \frac{u'_{\alpha\mu\mu\delta}}{d_{\alpha\mu\mu\delta}}\delta_{\beta\gamma} + \frac{u'_{\mu\beta\gamma\mu}}{d_{\mu\beta\gamma\mu}}\delta_{\alpha\delta} + \frac{u'_{\mu\beta\mu\delta}}{d_{\mu\beta\mu\delta}}\delta_{\alpha\gamma} + \frac{u'_{\mu\mu\gamma\delta}}{d_{\mu\mu\gamma\delta}}\delta_{\alpha\beta}\right)$$
$$-3\left(\frac{u'_{\mu\mu\nu\nu}}{d_{\mu\mu\nu\nu}}\delta_{\alpha\beta}\delta_{\gamma\delta} + \frac{u'_{\mu\nu\mu\nu}}{d_{\mu\nu\mu\nu}}\delta_{\alpha\gamma}\delta_{\beta\delta} + \frac{u'_{\mu\nu\nu\mu}}{d_{\mu\nu\nu\mu}}\delta_{\alpha\delta}\delta_{\beta\gamma}\right) \tag{A5c}$$

$$A_{\gamma,\alpha\beta} = -3\frac{u''_{\gamma,\alpha\beta}}{d_{\alpha\beta}} + u''_{\gamma,\mu\mu}\delta_{\alpha\beta} \tag{A5d}$$

$$B_{\gamma\delta,\alpha\beta} = -3\frac{u'''_{\gamma\delta,\alpha\beta}}{d_{\alpha\beta}} + u'''_{\gamma\delta,\mu\mu}\delta_{\alpha\beta} \tag{A5e}$$

$$C_{\alpha\beta,\gamma\delta} = -3\frac{u''_{\alpha\beta,\gamma\delta}}{d_{\alpha\beta}d_{\gamma\delta}} + \frac{u''_{\alpha\beta,\mu\mu}}{d_{\alpha\beta}}\delta_{\gamma\delta} + \frac{u''_{\mu\mu,\gamma\delta}}{d_{\gamma\delta}}\delta_{\alpha\beta} - \frac{1}{3}u''_{\mu\mu,\nu\nu}\delta_{\alpha\beta}\delta_{\gamma\delta} \tag{A5f}$$

$$D_{\delta,\alpha\beta\gamma} = -15\frac{u''_{\delta,\alpha\beta\gamma}}{d_{\alpha\beta\gamma}} + 3\left(\frac{u''_{\delta,\alpha\mu\mu}}{d_{\alpha\mu\mu}}\delta_{\beta\gamma} + \frac{u''_{\delta,\mu\beta\mu}}{d_{\mu\beta\mu}}\delta_{\alpha\gamma} + \frac{u''_{\delta,\mu\mu\gamma}}{d_{\mu\mu\gamma}}\delta_{\alpha\beta}\right) \qquad (A5g)$$

in which $d_{\alpha\beta}$ is the permutations of the subscripts $\alpha\beta$. Note that $\frac{u'_{\mu\mu}}{d_{\mu\mu}} = \frac{u'_{xx}}{d_{xx}} + \frac{u'_{yy}}{d_{yy}} + \frac{u'_{xx}}{d_{xx}} = u'_{\mu\mu}$ in the last expression of Eq. (A5a) since $d_{xx} = d_{yy} = d_{zz} = 1$, and similar for the last expressions of Eqs. (A5d–A5f). Eq. (A5) gives exactly the same traceless multipoles and multipole-multipole polarizabilities as in Eq. (20) of M&Y's derivation[12], as can be easily checked. Moreover, Eq. (A5) is equivalent to Eqs. (22) and (28), which give the multipoles and multipole-multipole polarizabilities in the Buckingham expansion in the traceless form, Eq. (27), and the permutation factor $d_{\alpha\beta\cdots}$ in Eq. (A5) is given by Eq. (A2), in which only one of the symmetry $d_{\alpha\beta\cdots}$ electric field gradient components is included. Of course, since Eq. (27) includes all the symmetric $E_{\alpha\beta\cdots}$, the division of $d_{\alpha\beta\cdots}$ is unavoidable for the conversion between Eq. (A3) and Eq. (12).

Finally, we note that some of the gaps in M&Y's derivation[12] was filled by Applequist[1984], who also adopted multi-variable Taylor series with all the electric field and gradient components included. The resultant expression is similar to Eq. (A3), but with all the Greek subscripts freely varying on $\{x, y, z\}$. Thus, Applequist's convention, with free permutations of subscripts and free exchanges of groups of subscripts, is also different than the Buckingham convention, and a conversion is needed. On the other hand, in the derivation we provide in Sec. 2 in this study, the convention is explicitly designed at very beginning. Thus, the derivation is naturally driven to the final form of Buckingham expansion in the traced form in Eq. (12), and no conversions are needed.

**Appendix B. Finite Field Method Based on Buckingham Expansion.**

Based on the discussions in Sec. 2, we propose the following general expression of Buckingham expansion in the traced form,

$$U = q\varphi - \sum_{n_\infty=0}^{N_\infty} \cdots \sum_{n_3=0}^{N_3} \sum_{n_2=0}^{N_2} \sum_{n_1=0}^{N_1} c_{n_1 n_2 n_3 \cdots} P_{n_1 n_2 n_3 \cdots} E_{n_1 n_2 n_3 \cdots} \tag{B1}$$

in which $P_{n_1 n_2 n_3 \cdots}$ denotes either multipole or (hyper)polarizability in the traced form, with $P_1 = P_{100\cdots} = \mu_\alpha$, $P_{01} = P_{0100\cdots} = Q_{\alpha\beta}$, $P_{11} = P_{1100\cdots} = a_{\gamma,\alpha\beta}$, $P_{02} = P_{0200\cdots} = c_{\alpha\beta,\gamma\delta}$, *etc.*, and $P_0 = P_{000\cdots} = -U_0$ as definition. It is understood that the exchange of the sub-indices, i.e., $P_{n_2 n_1 n_3 \cdots}$, is not allowed. For example, for $P_{11}$ with $n_1 = 1$, and $n_2 = 1$, so that $P_{n_1 n_2} = a_{\gamma,\alpha\beta}$ is allowed, but $P_{n_2 n_1} = a_{\alpha\beta,\gamma}$ is not allowed. On the other hand, the permutations of the Greek subscripts in the same place of $P_{n_1 n_2 n_3 \cdots}$ is allowed. For example, for $P_{02}$, both $c_{\alpha\beta,\gamma\delta}$ and $c_{\gamma\delta,\alpha\beta}$ are allowed because the groups of the Greek subscripts are inherently exchangeable, as discussed in Sec. 2. Also, the permutation of subscripts in the same group is allowed. For example, $a_{\gamma,\beta\alpha} = a_{\gamma,\alpha\beta}$ are allowed. Such convention is of course consistent with the Buckingham convention, as discussed in Sec. 2. In Eq. (B1),

$$c_{n_1 n_2 n_3 \cdots} = \frac{1}{n_1!} \frac{1}{2!^{n_2}} \frac{1}{3!^{n_3}} \cdots, \qquad E_{n_1 n_2 n_3 \cdots} = E_1^{n_1} E_2^{n_2} E_3^{n_3} \cdots \tag{B2}$$

where $E_l^{n_l}$ denotes $n_l$ power of the $l$th-rank electric field cartesian tensor, with $l$ Greek subscripts, so that $E_1^1 = E_\alpha$, $E_1^2 = E_\alpha E_\beta$, $E_2^1 = E_{\alpha\beta}$, $E_2^2 = E_{\alpha\beta} E_{\gamma\delta}$, and $E_l^0 = 1$, *etc.*.

Though Eq. (B1) can be expanded to $N_\infty$, for practical usage only a small portion is enough. For example, let $N_1 = 2$, $N_2 = 2$, and $N_l = 0$ for $l > 2$, we have the truncated version of Eq. (B1), i.e.,

$$U = q\varphi - \sum_{n_2=0}^{2}\sum_{n_1=0}^{2} c_{n_1 n_2} P_{n_1 n_2} E_{n_1 n_2}$$

$$= q\varphi - c_{00} P_{00} E_1^0 E_2^0 - c_{10} P_{10} E_1^1 E_2^0 - c_{20} P_{20} E_1^2 E_2^0 - c_{01} P_{01} E_1^0 E_2^1 - c_{11} P_{11} E_1^1 E_2^1$$

$$- c_{21} P_{21} E_1^2 E_2^1 - c_{02} P_{02} E_1^0 E_2^2 - c_{12} P_{12} E_1^1 E_2^2 - c_{22} P_{22} E_1^2 E_2^2$$

$$= q\varphi + U_0 - \mu_\alpha E_\alpha - \frac{1}{2}\alpha_{\alpha\beta} E_\alpha E_\beta - \frac{1}{2} Q_{\alpha\beta} E_{\alpha\beta} - \frac{1}{2} a_{\gamma,\alpha\beta} E_\gamma E_{\alpha\beta}$$

$$- \frac{1}{4} b_{\gamma\delta,\alpha\beta} E_\gamma E_\delta E_{\alpha\beta} - \frac{1}{4} c_{\alpha\beta,\gamma\delta} E_{\alpha\beta} E_{\gamma\delta} - \frac{1}{4} c^{(2)}_{\eta,\alpha\beta,\gamma\delta} E_\eta E_{\alpha\beta} E_{\gamma\delta} - \frac{1}{8} c^{(3)}_{\eta\xi,\alpha\beta,\gamma\delta} E_\eta E_\xi E_{\alpha\beta} E_{\gamma\delta}$$

which is comparable to Eq. (12), with an additional term $P_{22} = c^{(3)}_{\eta\xi,\alpha\beta,\gamma\delta}$ called dipole-dipole-quadrupole-quadrupole polarizability. Note that Einstein summation convention is applied on the repeated Greek subscripts in the above expression. For another example, when only the specific electric field $E_{n_1}$, i.e., ($\pm E_\alpha$, $\pm E_\beta$), is applied, the resultant Buckingham expansion is reduced to Taylor series of electric field, from which the finite difference expression of dipole polarizability $\alpha_{\alpha\beta}$ can be derived[25], as shown in Eq. (30) in Sec. 4.

We next derive the finite field expression of the traced quadrupole $Q_{\alpha\beta}$ using Eq. (B1). When only the specific electric field gradient of $E_{0n_2}$, i.e., $\pm E_{\alpha\beta}$, is applied, only the corresponding $P_{0n_2}$ responses to the power of $E_{\alpha\beta}^{n_2}$, so that Eq. (B1) can be simplified to include the contribution solely from $\pm E_{\alpha\beta}$, i.e.,

$$U(\pm E_{\alpha\beta}) = -\sum_{n_2=0}^{\infty} c_{0n_2} P_{0n_2} E_{0n_2}$$

$$= U_0 \mp \frac{1}{2} Q_{\alpha\beta} E_{\alpha\beta} - \frac{1}{4} c_{\alpha\beta,\alpha\beta} E_{\alpha\beta}^2 \mp \frac{1}{8} p_{\alpha\beta,\alpha\beta,\alpha\beta} E_{\alpha\beta}^3 - \frac{1}{16} p_{\alpha\beta,\alpha\beta,\alpha\beta,\alpha\beta} E_{\alpha\beta}^4 + O(E_2^5)$$

(B3)

in which $p_{\alpha\beta,\alpha\beta,\alpha\beta}$ is quadrupole-quadrupole-quadrupole polarizability, and $p_{\alpha\beta,\alpha\beta,\alpha\beta,\alpha\beta}$ is quadrupole-quadrupole-quadrupole-quadrupole polarizability, respectively. Note that Einstein summation convention is not applied on Eq. (B3) because the cartesian component $\alpha\beta$ is not dummy as it appears on the LHS of the expression. From Eq. (B3), $Q_{\alpha\beta}$ can be obtained by 2-side finite difference against $E_{\alpha\beta}$, i.e.,

$$\frac{U(-E_{\alpha\beta})-U(E_{\alpha\beta})}{E_{\alpha\beta}} = Q_{\alpha\beta} + \frac{1}{4} p_{\alpha\beta,\alpha\beta,\alpha\beta} E_{\alpha\beta}^2 + O(E_2^4) \qquad (B4a)$$

which gives $Q_{\alpha\beta}$ with accuracy up to $O(E_2^2)$. Doubling the electric field gradient, i.e., $\pm 2E_{\alpha\beta}$, we have

$$\frac{U(-2E_{\alpha\beta})-U(2E_{\alpha\beta})}{2E_{\alpha\beta}} = Q_{\alpha\beta} + p_{\alpha\beta,\alpha\beta,\alpha\beta} E_{\alpha\beta}^2 + O(E_2^4) \qquad (B4b)$$

Combining Eqs. (B4a) and (B4b) to eliminate the $p_{\alpha\beta,\alpha\beta,\alpha\beta} E_{\alpha\beta}^2$ term, we have

$$Q_{\alpha\beta} = \frac{4}{3} \times \frac{U(-E_{\alpha\beta})-U(E_{\alpha\beta})}{E_{\alpha\beta}} - \frac{1}{3} \times \frac{U(-2E_{\alpha\beta})-U(2E_{\alpha\beta})}{2E_{\alpha\beta}} + O(E_2^4) \qquad (B5)$$

The above expression is 4-side finite difference against $E_{\alpha\beta}$ with accuracy up to $O(E_2^4)$, and is adopted in Sec. 4 to calculate the traced quadrupole numerically.

Similarly, the quadrupole-quadrupole polarizability $c_{\alpha\beta,\gamma\delta}$ can be obtained by applying two specific electric field gradients, ($\pm E_{\alpha\beta}$, $\pm E_{\gamma\delta}$), simultaneously, and using Eq. (B1),

$$U(\pm E_{\alpha\beta}, E_{\gamma\delta}) = -\sum_{n_2=0}^{\infty} c_{0n_2} P_{0n_2} E_{0n_2}$$

$$= U_0 \mp \frac{1}{2} Q_{\alpha\beta} E_{\alpha\beta} - \frac{1}{2} Q_{\gamma\delta} E_{\gamma\delta} - \frac{1}{4} c_{\alpha\beta,\alpha\beta} E_{\alpha\beta}^2 \mp \frac{2}{4} c_{\alpha\beta,\gamma\delta} E_{\alpha\beta} E_{\gamma\delta} - \frac{1}{4} c_{\gamma\delta,\gamma\delta} E_{\gamma\delta}^2$$

$$\mp \frac{1}{8} p_{\alpha\beta,\alpha\beta,\alpha\beta} E_{\alpha\beta}^3 - \frac{3}{8} p_{\alpha\beta,\alpha\beta,\gamma\delta} E_{\alpha\beta}^2 E_{\gamma\delta} \mp \frac{3}{8} p_{\alpha\beta,\gamma\delta,\gamma\delta} E_{\alpha\beta} E_{\gamma\delta}^2 - \frac{1}{8} p_{\gamma\delta,\gamma\delta,\gamma\delta} E_{\gamma\delta}^3$$

$$- \frac{1}{16} p_{\alpha\beta,\alpha\beta,\alpha\beta,\alpha\beta} E_{\alpha\beta}^4 \mp \frac{4}{16} p_{\alpha\beta,\alpha\beta,\alpha\beta,\gamma\delta} E_{\alpha\beta}^3 E_{\gamma\delta} - \frac{6}{16} p_{\alpha\beta,\alpha\beta,\gamma\delta,\gamma\delta} E_{\alpha\beta}^2 E_{\gamma\delta}^2$$

$$\mp \frac{4}{16} p_{\alpha\beta,\gamma\delta,\gamma\delta,\gamma\delta} E_{\alpha\beta} E_{\gamma\delta}^3 - \frac{1}{16} p_{\gamma\delta,\gamma\delta,\gamma\delta,\gamma\delta} E_{\gamma\delta}^4$$

$$\mp \frac{1}{32} p_{\alpha\beta,\alpha\beta,\alpha\beta,\alpha\beta,\alpha\beta} E_{\alpha\beta}^5 - \frac{5}{32} p_{\alpha\beta,\alpha\beta,\alpha\beta,\alpha\beta,\gamma\delta} E_{\alpha\beta}^4 E_{\gamma\delta} \mp \frac{10}{32} p_{\alpha\beta,\alpha\beta,\alpha\beta,\gamma\delta,\gamma\delta} E_{\alpha\beta}^3 E_{\gamma\delta}^2$$

$$- \frac{10}{32} p_{\alpha\beta,\alpha\beta,\gamma\delta,\gamma\delta,\gamma\delta} E_{\alpha\beta}^2 E_{\gamma\delta}^3 \mp \frac{5}{32} p_{\alpha\beta,\gamma\delta,\gamma\delta,\gamma\delta,\gamma\delta} E_{\alpha\beta} E_{\gamma\delta}^4 - \frac{1}{32} p_{\gamma\delta,\gamma\delta,\gamma\delta,\gamma\delta,\gamma\delta} E_{\gamma\delta}^5 + O(E_2^6)$$

(B6a)

$$U(\pm E_{\alpha\beta}, -E_{\gamma\delta}) = -\sum_{n_2=0}^{\infty} c_{0n_2} P_{0n_2} E_{0n_2}$$

$$= U_0 \mp \frac{1}{2} Q_{\alpha\beta} E_{\alpha\beta} + \frac{1}{2} Q_{\gamma\delta} E_{\gamma\delta} - \frac{1}{4} c_{\alpha\beta,\alpha\beta} E_{\alpha\beta}^2 \pm \frac{2}{4} c_{\alpha\beta,\gamma\delta} E_{\alpha\beta} E_{\gamma\delta} - \frac{1}{4} c_{\gamma\delta,\gamma\delta} E_{\gamma\delta}^2$$

$$\mp \frac{1}{8} p_{\alpha\beta,\alpha\beta,\alpha\beta} E_{\alpha\beta}^3 + \frac{3}{8} p_{\alpha\beta,\alpha\beta,\gamma\delta} E_{\alpha\beta}^2 E_{\gamma\delta} \mp \frac{3}{8} p_{\alpha\beta,\gamma\delta,\gamma\delta} E_{\alpha\beta} E_{\gamma\delta}^2 + \frac{1}{8} p_{\gamma\delta,\gamma\delta,\gamma\delta} E_{\gamma\delta}^3$$

$$- \frac{1}{16} p_{\alpha\beta,\alpha\beta,\alpha\beta,\alpha\beta} E_{\alpha\beta}^4 \pm \frac{4}{16} p_{\alpha\beta,\alpha\beta,\alpha\beta,\gamma\delta} E_{\alpha\beta}^3 E_{\gamma\delta} - \frac{6}{16} p_{\alpha\beta,\alpha\beta,\gamma\delta,\gamma\delta} E_{\alpha\beta}^2 E_{\gamma\delta}^2$$

$$\pm \frac{4}{16} p_{\alpha\beta,\gamma\delta,\gamma\delta,\gamma\delta} E_{\alpha\beta} E_{\gamma\delta}^3 - \frac{1}{16} p_{\gamma\delta,\gamma\delta,\gamma\delta,\gamma\delta} E_{\gamma\delta}^4$$

$$\mp \frac{1}{32} p_{\alpha\beta,\alpha\beta,\alpha\beta,\alpha\beta,\alpha\beta} E_{\alpha\beta}^5 + \frac{5}{32} p_{\alpha\beta,\alpha\beta,\alpha\beta,\alpha\beta,\gamma\delta} E_{\alpha\beta}^4 E_{\gamma\delta} \mp \frac{10}{32} p_{\alpha\beta,\alpha\beta,\alpha\beta,\gamma\delta,\gamma\delta} E_{\alpha\beta}^3 E_{\gamma\delta}^2$$

$$+ \frac{10}{32} p_{\alpha\beta,\alpha\beta,\gamma\delta,\gamma\delta,\gamma\delta} E_{\alpha\beta}^2 E_{\gamma\delta}^3 \mp \frac{5}{32} p_{\alpha\beta,\gamma\delta,\gamma\delta,\gamma\delta,\gamma\delta} E_{\alpha\beta} E_{\gamma\delta}^4 + \frac{1}{32} p_{\gamma\delta,\gamma\delta,\gamma\delta,\gamma\delta,\gamma\delta} E_{\gamma\delta}^5 + O(E_2^6)$$

(B6b)

in which the coefficients in denominator are determined according to Eq. (B2), while the coefficients in numerator are determined according to permutation of the groups of the subscripts, which are inherently exchangeable, as discussed in Sec. 2. From Eq. (B6), $c_{\alpha\beta,\gamma\delta}$ can be obtained by 4-side finite difference against $E_{\alpha\beta}$ and $E_{\gamma\delta}$, i.e.,

$$\frac{U(-E_{\alpha\beta},E_{\gamma\delta})+U(E_{\alpha\beta},-E_{\gamma\delta})-U(E_{\alpha\beta},E_{\gamma\delta})-U(-E_{\alpha\beta},-E_{\gamma\delta})}{2E_{\alpha\beta}E_{\gamma\delta}} \tag{B7a}$$

$$= c_{\alpha\beta,\gamma\delta} + \frac{8}{16}p_{\alpha\beta,\alpha\beta,\alpha\beta,\gamma\delta}E_{\alpha\beta}^2 + \frac{8}{16}p_{\alpha\beta,\gamma\delta,\gamma\delta,\gamma\delta}E_{\gamma\delta}^2 + O(E_2^4)$$

which gives $c_{\alpha\beta,\gamma\delta}$ with accuracy up to $O(E_2^2)$. Doubling the electric field gradient, i.e., $\pm 2E_{\alpha\beta}$ and $\pm 2E_{\gamma\delta}$, we have

$$\frac{U(-2E_{\alpha\beta},2E_{\gamma\delta})+U(2E_{\alpha\beta},-2E_{\gamma\delta})-U(2E_{\alpha\beta},2E_{\gamma\delta})-U(-2E_{\alpha\beta},-2E_{\gamma\delta})}{8E_{\alpha\beta}E_{\gamma\delta}} \tag{B7b}$$

$$= c_{\alpha\beta,\gamma\delta} + \frac{32}{16}p_{\alpha\beta,\alpha\beta,\alpha\beta,\gamma\delta}E_{\alpha\beta}^2 + \frac{32}{16}p_{\alpha\beta,\gamma\delta,\gamma\delta,\gamma\delta}E_{\gamma\delta}^2 + O(E_2^4)$$

Combining Eqs. (B7a) and (B7b) to eliminate the $E_2^2$ terms, we have

$$c_{\alpha\beta,\gamma\delta} = \frac{4}{3}\times\frac{U(-E_{\alpha\beta},E_{\gamma\delta})+U(E_{\alpha\beta},-E_{\gamma\delta})-U(E_{\alpha\beta},E_{\gamma\delta})-U(-E_{\alpha\beta},-E_{\gamma\delta})}{2E_{\alpha\beta}E_{\gamma\delta}}$$

$$-\frac{1}{3}\times\frac{U(-2E_{\alpha\beta},2E_{\gamma\delta})+U(2E_{\alpha\beta},-2E_{\gamma\delta})-U(2E_{\alpha\beta},2E_{\gamma\delta})-U(-2E_{\alpha\beta},-2E_{\gamma\delta})}{8E_{\alpha\beta}E_{\gamma\delta}} + O(E_2^4)$$

(B8)

The above expression is 8-side finite difference against $E_{\alpha\beta}$ and $E_{\gamma\delta}$ with accuracy up to $O(E_2^4)$, and is adopted in Sec. 4 to calculate the traced quadrupole-quadrupole polarizability numerically. Note that the actual calculations for certain component may be simplified because the simultaneously applied two electric field gradients in opposite directions cancel each other. For example, using Eq. (B8), $c_{zz,zz}$ can be calculated by

$$c_{zz,zz} = \frac{4}{3}\times\frac{2U_0-U(2E_{zz})-U(-2E_{zz})}{2E_{zz}^2} - \frac{1}{3}\times\frac{2U_0-U(4E_{zz})-U(-4E_{zz})}{8E_{zz}^2} + O(E_{zz}^4)$$

in which $U_0$ is the energy with zero external gradient because the simultaneously applied ($+E_{zz}$, $-E_{zz}$) cancel each other. Therefore, only 5, instead of 8, *ab initio* calculations are needed to calculate $c_{zz,zz}$ numerically with finite field method.

The above derivation on $Q_{\alpha\beta}$ and $c_{\alpha\beta,\gamma\delta}$ are general and can be easily applied to calculate other traced multipoles or (hyper)polarizabilities of interest in the frame of finite field method. To obtain accurate polarizabilities/hyperpolarizabilities, wavefunction-based methods such as CCSD(T) using the finite field approach are necessary. And for large π-conjugated systems such as polydiacetylene(PDA) and polybutatriene(PBT) as shown in previous works[30, 31], DFT methods with improved exchange-correlation(XC) functionals are also good choices, when considering the balance between accuracy and computational cost. It should be noted that our finite field method derived in appendix B could be applied in conjunction with both wavefunction-based and DFT methods to getting higher accuracy. The transformations to the corresponding traceless multipoles or (hyper)polarizabilities are discussed in Sec. 3 as well as Appendix C.

**Appendix C. General Buckingham Expansion in the traceless form.**

The general formula of the electric field gradients is [13, 32]

$$\begin{aligned}
E_M &= E_{\alpha_1\alpha_2\cdots\alpha_M} \\
&= eR^{-(2M+1)} \sum_{L=0}^{\left[\frac{M}{2}\right]} (-1)^{M-L} \left[2(M-L)-1\right]!! R^{2L} \left(R^{M-2L}\delta^L\right)_{\alpha_1\alpha_2\cdots\alpha_M} \quad (C1) \\
&= \sum_{L=0}^{\left[\frac{M}{2}\right]} (-1)^L \frac{\left[2(M-L)-1\right]!!}{(2M-1)!!} \left(F\delta^L\right)_{\alpha_1\alpha_2\cdots\alpha_M}
\end{aligned}$$

in which $M$ stands for the rank of the traceless electric field gradients $E_M = E_{\alpha_1\alpha_2\cdots\alpha_M}$, $F = F_M = F_{\alpha_1\alpha_2\cdots\alpha_M}$ denotes the corresponding auxiliary traced electric field gradients, $[M/2]$ represents the largest integer not exceeding $M/2$. And in Eq. (C1),

$$\left(R^{M-2L}\delta^L\right)_{\alpha_1\alpha_2\cdots\alpha_M}=\sum\delta_{\alpha_1\alpha_2}\delta_{\alpha_3\alpha_4}\cdots\delta_{\alpha_{2L-1}\alpha_{2L}}R_{\alpha_{2L+1}}R_{\alpha_{2L+2}}\cdots R_{\alpha_{M-1}}R_{\alpha_M}$$
$$\left(F\delta^L\right)_{\alpha_1\alpha_2\cdots\alpha_M}=\sum\delta_{\alpha_1\alpha_2}\delta_{\alpha_3\alpha_4}\cdots\delta_{\alpha_{2L-1}\alpha_{2L}}F_{\mu\mu\nu\nu\cdots\xi\xi\alpha_{2L+1}\alpha_{2L+2}\cdots\alpha_{M-1}\alpha_M}$$
(C2)

in which the sum is over all the different permutations of the subscripts $\alpha_1, \alpha_2, \cdots, \alpha_M$, including $N_{M,L} = M!/[2^L L!(M-2L)!]$ terms.

Based on the Eqs. (C1) and (C2), the general formula of the Eqs. (21a-21c) is given as

$$J_{E_M,F_M}=J_{\alpha_1\alpha_2\cdots\alpha_M,\beta_1\beta_2\cdots\beta_M}$$
$$=\frac{\partial E_{\alpha_1\alpha_2\cdots\alpha_M}}{\partial F_{\beta_1\beta_2\cdots\beta_M}}=\sum_{L=0}^{\left[\frac{M}{2}\right]}(-1)^L\frac{[2(M-L)-1]!!}{(2M-1)!!}\left(\delta^{M-2L}_{\alpha_o\beta_o}\delta^L_{\alpha_p\alpha_q}\delta^L_{\beta_p\beta_q}\right)_{\substack{\alpha_1\alpha_2\cdots\alpha_M \\ \beta_1\beta_2\cdots\beta_M}} \quad (C3)$$

and

$$\left(\delta^{M-2L}_{\alpha_o\beta_o}\delta^L_{\alpha_p\alpha_q}\delta^L_{\beta_p\beta_q}\right)_{\substack{\alpha_1\alpha_2\cdots\alpha_M \\ \beta_1\beta_2\cdots\beta_M}}=\sum\delta_{\alpha_1\alpha_2}\delta_{\beta_1\beta_2}\delta_{\alpha_3\alpha_4}\delta_{\beta_3\beta_4}\cdots\delta_{\alpha_{2L-1}\alpha_{2L}}\delta_{\beta_{2L-1}\beta_{2L}}\delta_{\alpha_{2L+1}\beta_{2L+1}}\delta_{\alpha_{2L+2}\beta_{2L+2}}\cdots\delta_{\alpha_M\beta_M}$$
(C4)

in which the sum is over the all the different permutations of the subscripts $\alpha_1, \alpha_2, \cdots, \alpha_M$, and the permutations of the subscripts $\beta_1, \beta_2, \cdots, \beta_M$ need to follow the same order.

With the help of the Eqs. (C3) and (C4), the general formula from traced to traceless multipoles and (hyper)polarizabilities is given as

$$\Xi_{n_1n_2n_3\cdots}=\mathfrak{I}_{n_1n_2n_3\cdots}P_{n_1n_2n_3\cdots} \quad (C5)$$

in which $P_{n_1n_2n_3\cdots}$ is the traced multipole or (hyper)polarizability defined in the Eq. (B1), and $\Xi_{n_1n_2n_3\cdots}$ denotes the corresponding traceless multipole or

(hyper)polarizability, with $\Xi_{01} = \Xi_{0100\cdots} = \Theta_{\alpha\beta}$, $\Xi_{11} = \Xi_{1100\cdots} = A_{\gamma,\alpha\beta}$, $\Xi_{02} = \Xi_{0200\cdots} = C_{\alpha\beta,\gamma\delta}$, etc., while $\Xi_1 = \Xi_{100\cdots} = P_{100\cdots} = \mu_\alpha$, $\Xi_2 = \Xi_{200\cdots} = P_{200\cdots} = \alpha_{\alpha\beta}$, $\Xi_3 = \Xi_{300\cdots} = P_{300\cdots} = \beta_{\alpha\beta\gamma}$, $\Xi_4 = \Xi_{400\cdots} = P_{400\cdots} = \gamma_{\alpha\beta\gamma\delta}$, and $\Xi_0 = \Xi_{000\cdots} = P_{000\cdots} = -U_0$ as definition. In Eq. (C5), the operator $\mathfrak{I}_{n_1 n_2 n_3 \cdots}$ is

$$\mathfrak{I}_{n_1 n_2 n_3 \cdots} = \prod_M^{\{M|n_M \neq 0, M=2,3,\cdots\}} f_M \left(J_{E_M, F_M}\right)^{n_M}$$
$$= \prod_M^{\{M|n_M \neq 0, M=2,3,\cdots\}} \frac{(2M-1)!!}{M!} \left(J_{E_M, F_M}\right)^{n_M} \quad (C6)$$

in which just one corresponding $f_M$ factor in the Eq. (23) is needed in the operator when $n_M \neq 0$, while $J_{E_M, F_M}$ need to be carried out $n_M$ times to satisfy the traceless conditions.

Based on the discussions in Sec. 3 and the Eqs. (C1-C6), we propose the following general expression of Buckingham expansion in the traceless form,

$$U = q\varphi - \sum_{n_\infty=0}^{N_\infty} \cdots \sum_{n_3=0}^{N_3} \sum_{n_2=0}^{N_2} \sum_{n_1=0}^{N_1} f_{n_1 n_2 n_3 \cdots} c_{n_1 n_2 n_3 \cdots} \Xi_{n_1 n_2 n_3 \cdots} E_{n_1 n_2 n_3 \cdots} \quad (C7)$$

in which $c_{n_1 n_2 n_3 \cdots}$ and $E_{n_1 n_2 n_3 \cdots}$ have been defined in Eq. (B2). And in Eq. (C7),

$$f_{n_1 n_2 n_3 \cdots} = \prod_M^{\{M|n_M \neq 0, M=2,3,\cdots\}} \frac{M!}{(2M-1)!!} = \prod_M^{\{M|n_M \neq 0, M=2,3,\cdots\}} \frac{1}{f_M} \quad (C8)$$

where $f_{n_1 n_2 n_3 \cdots}$ is the conversion coefficient from traced to traceless multipole or multipole-multipole polarizability tensor, defining that when $n_2 = n_3 = \cdots = 0$, $f_{n_1 n_2 n_3 \cdots} = 1$.

It is understood that in the Eq. (C7), the exchange of the sub-indices, i.e., $\Xi_{n_2 n_1 n_3 \ldots}$, is not allowed. For example, for $\Xi_{11}$ with $n_1 = 1$, and $n_2 = 1$, so that $\Xi_{n_1 n_2} = A_{\gamma,\alpha\beta}$ is allowed, but $\Xi_{n_2 n_1} = A_{\alpha\beta,\gamma}$ is not allowed. On the other hand, the permutations of the Greek subscripts in the same place of $\Xi_{n_1 n_2 n_3 \ldots}$ is allowed. For example, for $\Xi_{02}$, both $C_{\alpha\beta,\gamma\delta}$ and $C_{\gamma\delta,\alpha\beta}$ are allowed because the groups of the Greek subscripts are inherently exchangeable, as discussed in the text. Also, the permutation of subscripts in the same group is allowed. For example, $A_{\gamma,\beta\alpha} = A_{\gamma,\alpha\beta}$ are allowed. Such convention is consistent with the Buckingham convention, as discussed in the text. It is noted that the Eq. (C8) means that when $n_M \neq 0$, just one corresponding $1/f_M$ factor is needed in the transform, as shown in Eqs. (26a-26c). For example, when $n_1 = 0$, $n_2 = 2$, the corresponding energy term transform in Eq. (26a) from traced to traceless expression is $-\frac{1}{4} c_{\alpha\beta,\gamma\delta} E_{\gamma\delta} E_{\alpha\beta} = -\frac{1}{f_\Theta} \frac{1}{4} f_\Theta c_{\alpha\beta,\gamma\delta} E_{\gamma\delta} E_{\alpha\beta} = -\frac{1}{6} C_{\alpha\beta,\gamma\delta} E_{\gamma\delta} E_{\alpha\beta}$, in which $f_\Theta$ appears just once.

Though Eq. (C7) can be expanded to $N_\infty$, for practice only a small portion is enough. For example, let $N_1 = 2$, $N_2 = 2$, and $N_l = 0$ for $l > 2$, we have the truncated version of Eq. (C7), i.e.,

$$U = q\varphi - \sum_{n_2=0}^{2} \sum_{n_1=0}^{2} f_{n_1 n_2} c_{n_1 n_2} \Xi_{n_1 n_2} E_{n_1 n_2}$$
$$= q\varphi - f_{00} c_{00} \Xi_{00} E_1^0 E_2^0 - f_{10} c_{10} \Xi_{10} E_1^1 E_2^0 - f_{20} c_{20} \Xi_{20} E_1^2 E_2^0 - f_{01} c_{01} \Xi_{01} E_1^0 E_2^1 - f_{11} c_{11} \Xi_{11} E_1^1 E_2^1$$
$$- f_{21} c_{21} \Xi_{21} E_1^2 E_2^1 - f_{02} c_{02} \Xi_{02} E_1^0 E_2^2 - f_{12} c_{12} \Xi_{12} E_1^1 E_2^2 - f_{22} c_{22} \Xi_{22} E_1^2 E_2^2$$
$$= q\varphi + U_0 - \mu_\alpha E_\alpha - \frac{1}{2} \alpha_{\alpha\beta} E_\alpha E_\beta - \frac{1}{3} \Theta_{\alpha\beta} E_{\alpha\beta} - \frac{1}{3} A_{\gamma,\alpha\beta} E_\gamma E_{\alpha\beta}$$
$$- \frac{1}{6} B_{\gamma\delta,\alpha\beta} E_\gamma E_\delta E_{\alpha\beta} - \frac{1}{6} C_{\alpha\beta,\gamma\delta} E_{\alpha\beta} E_{\gamma\delta} - \frac{1}{6} C^{(2)}_{\eta,\alpha\beta,\gamma\delta} E_\eta E_{\alpha\beta} E_{\gamma\delta} - \frac{1}{12} C^{(3)}_{\eta\xi,\alpha\beta,\gamma\delta} E_\eta E_\xi E_{\alpha\beta} E_{\gamma\delta}$$

which is consistent with the Eq. (27).

Other conventions may be adopted in the energy expansion besides Buckingham convention discussed above. For example, Applequist[13] defined the general traced multipoles or (hyper)polarizabilities as

$$p^{(k_1,k_2,\cdots,k_t)} = -\left(\frac{\partial^t U}{\partial E^{(k_1)} \partial E^{(k_2)} \cdots \partial E^{(k_t)}}\right) \tag{C9}$$

in which $U$ is the total electrostatic energy, $E^{(k_i)}$ represents electric field or gradients of rank $k_t$ which is the same as our traceless definition, i.e., Eq. (C1) and when $t = 1$, $p^{(k_1)}$ is the traced multipole with $p^{(1)} = \mu$, $p^{(2)} = 1/2Q$, $p^{(3)} = 1/6O$, $p^{(4)} = 1/24H$, $\cdots$. Applequist expressed his energy expansion as

$$U = q\varphi - \sum_{t=1}^{\infty} \frac{1}{t!} \sum_{k_1,k_2,\cdots,k_t} p^{(k_1,k_2,\cdots,k_t)} E^{(k_1)} E^{(k_2)} \cdots E^{(k_t)} \tag{C10}$$

in which $k_1$, $k_2$, $\cdots$, $k_t$ could traverse from 1 to $\infty$. The Eq. (C10) is just the multi-variable Taylor expansion where the variables are the electric field and gradients $E^{(k)}$. The convention adopted by Applequist in the Eq. (C10) is that the permutations of groups $k_1$, $k_2$, $\cdots$, $k_t$ are permitted which is different from Buckingham convention, and the permutations of subscripts inside the group are also permitted which is the same as Buckingham convention. For example, there exists both $p^{(1,2)}$ and $p^{(2,1)}$ terms in the Eq. (C10), while there just exists $a_{\gamma,\alpha\beta}$, not $a_{\alpha\beta,\gamma}$ in the Eq. (B1). It should be noted that the energy in the Eqs. (B1) and (C10) are equal. By comparing the Eqs. (B1) and (C10), we could find the relationships between Applequist's and our traced definition, i.e.,

$$p^{(k_1,k_2,\cdots,k_t)} = \prod_{M}^{\{M|n_M \neq 0, M=2,3,\cdots\}} \frac{n_M!}{M!^{n_M}} P_{n_1 n_2 n_3 \cdots} \tag{C11}$$

in which $P_{n_1 n_2 n_3 \cdots}$ is the traced multipole or (hyper)polarizability defined in the Eq. (B1) and $n_1 + n_2 + n_3 + \cdots = t$. Then Applequist defined the general traceless multipole and (hyper)polarizabilities as

$$\xi^{(k_1,k_2,\cdots,k_t)} = T_{k_1} T_{k_2} \cdots T_{k_t} \boldsymbol{p}^{(k_1,k_2,\cdots,k_t)} \tag{C12}$$

in which $T_{k_i} = (2k_i - 1)!! J_{E_{k_i}, F_{k_i}}$ and the operator $T_{k_i}$ just acts on those indices of the corresponding rank $k_t$ of $\boldsymbol{p}^{(k_1,k_2,\cdots,k_t)}$. And when $t = 1$, $\xi^{(k_1)}$ is the traceless multipole with $\xi^{(1)} = \boldsymbol{\mu}$, $\xi^{(2)} = \boldsymbol{\Theta}$, $\xi^{(3)} = \boldsymbol{\Omega}$, $\xi^{(4)} = \boldsymbol{\Phi}$, $\cdots$. With the help of the Eqs. (C5), (C6), (C11), and (C12), the relationship between Applequist's and Buckingham's traceless multipoles or (hyper)polarizbilities is

$$\begin{aligned}
\xi^{(k_1,k_2,\cdots,k_t)} &= \prod_M^{\{M | n_M \neq 0, M = 2,3,\cdots\}} n_M! (f_M)^{n_M - 1} \, \Xi_{n_1 n_2 n_3 \cdots} \\
&= \prod_M^{\{M | n_M \neq 0, M = 2,3,\cdots\}} n_M! \left[\frac{(2M-1)!!}{M!}\right]^{n_M - 1} \Xi_{n_1 n_2 n_3 \cdots}
\end{aligned} \tag{C13}$$

in which $\Xi_{n_1 n_2 n_3 \cdots}$ is the traceless multipole or (hyper)polarizbility defined in the Eq. (C5) and $n_1 + n_2 + n_3 + \cdots = t$.

**References**

[1] Buckingham AD, Fowler PW, Hutson JM. Theoretical studies of van der Waals molecules and intermolecular forces. *Chem Rev*, 1988, 88(6): 963-988
[2] Stone AJ. *The Theory of Intermolecular Forces*. Oxford University Press, 2013
[3] Buckingham AD. Direct Method of Measuring Molecular Quadrupole Moments. *J Chem Phys*, 1959, 30(6): 1580-1585
[4] Buckingham AD. Molecular quadrupole moments. *Quart Rev Chem Soc*, 1959, 13(3): 183-214
[5] Buckingham AD. Permanent and Induced Molecular Moments and Long-Range lntermolecular Forces. *Adv Chem Phys*, 1967, 12: 107-142
[6] Buckingham AD, Pople JA. Theoretical Studies of the Kerr Effect I: Deviations from a Linear Polarization Law. *Proc Phys Soc A*, 1955, 68: 905-909
[7] Buckingham AD, Disch RL. The quadrupole moment of the carbon dioxide molecule. *Proc Roy Soc A*, 1963, 273(1353): 275-289

[8] Bishop DM, Pipin J. Field and field-gradient polarizabilities of $H_2O$. *Theor Chim Acta*, 1987, 71(4): 247-253

[9] Maroulis G. Hyperpolarizability of $H_2O$ revisited: accurate estimate of the basis set limit and the size of electron correlation effects. *Chem Phys Lett*, 1998, 289(3-4): 403-411

[10] Haskopoulos A, Maroulis G. Dipole and quadrupole (hyper) polarizability for the asymmetric stretching of carbon dioxide: Improved agreement between theory and experiment. *Chem Phys Lett*, 2006, 417(1-3): 235-240

[11] Elking DM, Perera L, Duke R, Darden T, Pedersen LG. A Finite Field Method for Calculating Molecular Polarizability Tensors for Arbitrary Multipole Rank. *J Comput Chem*, 2011, 32(15): 3283-3295

[12] McLean AD, Yoshimine M. Theory of molecular polarizabilities. *J Chem Phys*, 1967, 47(6): 1927-1935

[13] Applequist J. Fundamental relationships in the theory of electric multipole moments and multipole polarizabilities in static fields. *Chem Phys*, 1984, 85(2): 279-290

[14] Thakkar AJ, Lupinetti C. Atomic polarizabilities and hyperpolarizabilities: A critical compilation. In: Maroulis G, Eds. Atoms, Molecules and Clusters in Electric Fields: Theoretical Approaches to the Calculation of Electric Polarizability. Imperial College Press, Lodon, 2006

[15] Frisch MJ, Trucks GW, Schlegel HB, Scuseria GE, Robb MA, Cheeseman JR, Scalmani G, Barone V, Petersson GA, Nakatsuji H, Li X, Caricato M, Marenich AV, Bloino J, Janesko BG, Gomperts R, Mennucci B, Hratchian HP, Ortiz JV, Izmaylov AF, Sonnenberg JL, Williams-Young D, Ding F, Lipparini F, Egidi F, Goings J, Peng B, Petrone A, Henderson T, Ranasinghe D, Zakrzewski VG, Gao J, Rega N, Zheng G, Liang W, Hada M, Ehara M, Toyota K, Fukuda R, Hasegawa J, Ishida M, Nakajima T, Honda Y, Kitao O, Nakai H, Vreven T, Throssell K, Montgomery JAJ, Peralta JE, Ogliaro F, Bearpark MJ, Heyd JJ, Brothers EN, Kudin KN, Staroverov VN, Keith TA, Kobayashi R, Normand J, Raghavachari K, Rendell AP, Burant JC, Iyengar SS, Tomasi J, Cossi M, Millam JM, Klene M, Adamo C, Cammi R, Ochterski JW, Martin RL, Morokuma K, Farkas O, Foresman JB, Fox DJ. Gaussian 16, Revision B.01. Gaussian, Inc., Wallingford CT., 2016

[16] Rhoderick EH. Definition of Nuclear Quadrupole Moments. *Nature*, 1947, 160(4060): 255-256

[17] Chetty N, Couling VW. Measurement of the electric quadrupole moments of $CO_2$ and OCS. *Mol Phys*, 2011, 109(5): 655-666

[18] Watson JN, Craven IE, Ritchie GLD. Temperature dependence of electric field-gradient induced birefringence in carbon dioxide and carbon disulfide. *Chem Phys Lett*, 1997, 274(1-3): 1-6

[19] Graham C, Imrie DA, Raab RE. Measurement of the electric quadrupole moments of $CO_2$, CO, $N_2$, $Cl_2$ and $BF_3$. *Mol Phys*, 1998, 93(1): 49-56

[20] Balachandran Pillai PC, Couling VW. Dispersion of the Rayleigh light-scattering virial coefficients and polarisability anisotropy of $CO_2$. *Mol Phys*, 2019, 117(3): 289-297

[21] Gentle IR, Laver DR, Ritchie GLD. Second hyperpolarizability and static polarizability anisotropy of carbon dioxide. *J Phys Chem*, 1989, 93(8): 3035-3038

[22] Maroulis G, Thakkar AJ. Polarizabilities and hyperpolarizabilities of carbon dioxide. *J Chem Phys*, 1990, 93(6): 4164-4171

[23] Maroulis G. Electric (hyper) polarizability derivatives for the symmetric stretching of carbon dioxide. *Chem Phys*, 2003, 291(1): 81-95

[24] Graner G, Rossetti C, Bailly D. The carbon dioxide molecule: A test case for the $r_0$, $r_e$ and $r_m$ structures. *Mol Phys*, 1986, 58(3): 627-636

[25] Kurtz HA, Stewart JJ, Dieter KM. Calculation of the nonlinear optical properties of molecules. *J Comput Chem*, 1990, 11(1): 82-87

[26] Alms GR, Burnham AK, Flygare WH. Measurement of the dispersion in polarizability anisotropies. *J Chem Phys*, 1975, 63(8): 3321-3326

[27] Bogaard MP, Buckingham AD, Pierens RK, White AH. Rayleigh scattering depolarization ratio and molecular polarizability anisotropy for gases. *J Chem Soc Faraday Trans 1*, 1978, 74: 3008-3015

[28] Maroulis G. A note on the electric multipole moments of carbon dioxide. *Chem Phys Lett*, 2004, 396(1): 66-68

[29] Loboda O, Ingrosso F, Ruiz-López MF, Reis H, Millot C. Dipole and Quadrupole Polarizabilities of the Water Molecule as a Function of Geometry. *J Comput Chem*, 2016, 37(23): 2125-2132

[30] Nénon S, Champagne B, Spassova MI. Assessing long-range corrected functionals with physically-adjusted range-separated parameters for calculating the polarizability and the second hyperpolarizability of polydiacetylene and polybutatriene chains. *Phys Chem Chem Phys*, 2014, 16(15): 7083-7088

[31] Oviedo MB, Ilawe NV, Wong BM. Polarizabilities of π-conjugated chains revisited: improved results from broken-symmetry range-separated DFT and new CCSD(T) benchmarks. *J Chem Theory Comput*, 2016, 12(8): 3593-3602

[32] Burgos E, Bonadeo H. Electrical multipoles and multipole interactions: compact expressions and a diagrammatic method. *Mol Phys*, 1981, 44(1): 1-15